\documentclass[pra, twocolumn,showpacs,nofootinbibfloatfix,amsmath,amsfonts,amssymb]{revtex4-1}%
\usepackage{amsmath,amsfonts,amssymb,color}
\usepackage{amsthm}
\usepackage{leftidx}
\usepackage{graphicx}
\usepackage{xcolor}
\usepackage{dcolumn}
\usepackage{bm}
\usepackage{epstopdf}
\usepackage{epsfig}
\newcommand{\n}{\noindent}

\begin{document}
	\title{Quantum computation via Floquet topological edge modes}
	\author{Raditya Weda Bomantara}
	\email{phyrwb@nus.edu.sg}
	\affiliation{%
		Department of Physics, National University of Singapore, Singapore 117543
	}
	\author{Jiangbin Gong}%
	\email{phygj@nus.edu.sg}
	\affiliation{%
		Department of Physics, National University of Singapore, Singapore 117543
	}
	\date{\today}
	
	%%%%%%%%%%%%%%%%%%%% ABSTRACT %%%%%%%%%%%%%%%%%%%%%%%%
	%\begin{linenumbers}
	
	\vspace{2cm}
	
	\begin{abstract}

Floquet topological matter has emerged as one exciting platform to explore rich physics and game-changing applications of topological phases.  As one remarkable and recently discovered feature of Floquet symmetry protected topological (SPT) phases, in principle a simple periodically driven system can host an arbitrary number of topological protected zero edge modes and $\pi$ edge modes, with Majorana zero modes and Majorana $\pi$ modes as examples protected by the particle-hole symmetry.  This work advocates a new route to holonomic quantum computation by exploiting the co-existence of many Floquet SPT edge modes, all of which have trivial dynamical phases during a computation protocol. As compelling evidence supporting this ambitious goal, three pairs of Majorana edge modes, hosted by a periodically driven one-dimensional (1D) superconducting superlattice, are shown to suffice to encode two logical qubits, realize quantum gate operations, and execute two simple quantum algorithms through adiabatic lattice deformation.  When compared with early studies on quantum computation based on Majorana zero modes of topological quantum wires,  significant resource saving is now made possible by use of Floquet SPT phases.  %This paper is thus hoped to motivate a series of future studies on the potential of Floquet topological matter in quantum computation.

	\end{abstract}
	%\pacs{03.65.Vf, 05.60.Gg, 05.30.Rt, 73.20.At}
	
	\maketitle
	
\section{Introduction} 	

Fault-tolerant quantum computation has been sought as a long term goal towards the development of quantum computers. Potential candidates for this purpose are Majorana zero modes (MZMs) emerging at the vortices or edges of topological superconductors \cite{Kit,ising,Ivanov}, which possess topological protection at the hardware level. In such systems, a qubit is encoded nonlocally from a pair of MZMs separated far apart from each other, and quantum gate operations are achieved by braiding them around each other \cite{tqc2}. Due to the constraints put in place by fermionic parity conservation and the number of MZMs that can be generated in a given system, Majorana-based quantum computation usually requires intricate geometry \cite{Tjun1,Tjun2,Tjun3,Tjun4} to initialize qubits and facilitate braiding between a pair of MZMs,  posing some difficulties in scaling it up to solve heavy computational tasks.

It is therefore of fundamental interest to seek innovative and alternative quantum computation schemes with considerable error tolerance on the hardware level.  In this work
we advocate to exploit an unusual feature of the so-called Floquet topological matter to realize holonomic quantum computation.
In recent years Floquet topological matter has emerged as one exciting platform to explore rich physics and potentially game-changing applications of topological phases.  In periodically driven systems, energy is no longer a conserved quantity and is replaced by the so-called quasienergy which is only defined modulo $2\pi/T$, with $T$ the driving period. As one recently discovered feature of Floquet symmetry protected topological (SPT) phases, in principle a simple periodically driven system can host an arbitrary number of topological zero edge modes and $\pi$ edge modes (with quasi-energy $0$ and $\pi/T$ respectively) \cite{Derek1,Longwen1,Longwen2}, with MZMs and Majorana $\pi$ modes as examples in the presence of the particle-hole symmetry \cite{opt,kk3,kk1,kk2,RG}. Given that both MZMs and Majorana $\pi$ modes yield a zero dynamical phase at even multiples of $T$, their coexistence presents a motivating case in reconsidering holonomic quantum computation. As to other parts of a system not directly hosting the edge modes, they can be deemed as auxiliary components, necessary to ensure topology-based fault tolerance inherent in the edge modes and also serving as  temporary information storage.

To advocate such a promising marriage between Floquet topological matter and quantum computation, one naturally starts with a one-dimensional (1D) prototype system capable of hosting multiple Floquet Majorana modes.
One also hopes that these Floquet Majorana modes are manipulable in order to accomplish braiding between them and consequently quantum gate operations without the need of introducing branched geometries of a quantum wire.  To this end, our previous study \cite{RG} has moved the first encouraging step by considering a periodically driven topological superconducting wire.
 %This allows additional Majorana modes to emerge at quasienergy $\pi$ \cite{opt,kk1,kk2,RG},
% which can then be braided with the already existing Majorana zero modes to encode and manipulate qubits \cite{RG}.
In particular,  though a 1D static topological superconductor typically hosts only a single MZM at each end, the application of periodic driving can add another pair of Majorana $\pi$ edge modes, thus yielding the minimal number of Majorana modes required to encode a single qubit \cite{RG}. Braiding between the Floquet MZM and Floquet $\pi$ mode therein and hence single-qubit gate operations were indeed shown to be feasible by using adiabatic lattice deformation alone. It thus becomes necessary and significant to explore the full potential of the coexistence of multiple or even many Floquet topological edge modes hosted by one single quantum wire.

Models with the particle-hole symmetry such as the Kitaev model \cite{Kit} naturally hosts MZMs and a periodically driven version may add the Majorana $\pi$ modes. SPT edge modes due to other symmetries are also of great interest \cite{SPT}, but are not directly useful for topologically protected quantum computation. Take, for example, the edge modes in the 1D Su-Schrieffer-Heeger (SSH) \cite{SSH} model. Despite that SSH edge modes are also pinned at zero energy, they are protected by the chiral instead of particle-hole symmetry. Because each SSH edge mode is already fermionic (rather than half-fermionic) in nature, one cannot combine two such edge modes to form a qubit sharing the same feature of Majorana qubits.  Nevertheless, the other side of the story is stimulating.  That is, a single SSH-like edge mode can be broken down into two Majorana fermions, each of which carries zero energy and is thus an MZM. The same philosophy applies to SSH-like $\pi$ edge modes.  This being the case, a single SSH-like edge mode afforded by the chiral symmetry can be used to encode a (local) qubit.  Two such edge modes localized at two opposite ends of a 1D wire are however far apart and their MZM constituents cannot be braided.

Given our general insights above, we construct a working model here with a periodically driven superconducting superlattice with both chiral and particle-hole symmetries. In the absence of periodic driving, such type of quantum wires can host either SSH- or  Kitaev-like edge modes \cite{SSHK1,SSHK2}(that is, SSH- or  Kitaev-like edge modes cannot coexist for any given set of system parameters). In the presence of periodic driving, it becomes possible for the SSH- and Kitaev-like edge modes to coexist, one of which is pinned at quasienergy zero, whereas the other is pinned at quasienergy $\pi/T$.  Intriguing quantum gate operations can then be anticipated. Indeed, one pair of SSH-like edge modes, viewed as two pairs of constituent Majorana modes, can now be exploited for information encoding and gate operations because of the possibility of braiding one of the Majorana constituents of a SSH-like edge mode with the other isolated Kitaev-like edge mode at quasienergy $\pi/T$.

To demonstrate the feasibility of the quantum computation scheme outline above, we restrict ourselves to the situation where in total three pairs of Majorana edge modes are hosted by a driven quantum wire.
After taking into account the fermionic parity conservation, two logical qubits can be constructed. { This work can thus be considered as an extension of Ref.~\cite{RG}, where only a single qubit was obtained with Kitaev-like edge states}.
%Because more Majorana modes are essentially clustered together at each end of the system, different pairs of %Majorana modes can be readily braided, allowing more gate operations to be accessible in our system.
In addition to the explicit construction of the logical qubits based on Floquet topological edge modes,
the protocols to accomplish the braiding between different pairs of Majorana modes are one main focus of this paper. We outline a proposal to readout the qubits by breaking the system's chiral symmetry. We also  demonstrate how our quantum computation scheme can be applied to implement two simple quantum algorithms with one single quantum wire.  At the end of this work we also discuss how to scale up our quantum computation scheme by explicitly showing how controlled-not (CNOT) gates can be realized with the use of two quantum wires.

This paper is structured as follows. We start in Sec.~\ref{background1} with a short review of Floquet theory to describe time-periodic (Floquet) systems and discuss the emergence of symmetry protected topological edge modes at quasienergy zero and $\pi/T$. In Sec.~\ref{holonomy}, we adapt the theory of adiabatic processes and holonomy to Floquet systems. We present our model in Sec.~\ref{models}, along with its symmetry properties and $Z\times Z$ topological invariants characterizing the emergence of zero and $\pi$ edge modes. In Sec.~\ref{encoding}, we show how the two different species of zero and $\pi$ edge modes can be written in terms of Majorana operators, which can in turn be used to encode two qubits. In Sec.~\ref{compute}, we present the application of such edge modes in holonomic quantum computation. In particular, we explicitly develop protocols to realize various single-gate operations by adiabatically deforming the system's Hamiltonian in various closed cycles, propose a means to readout qubits, demonstrate the implementation of two simple quantum algorithms with our system, and discuss the possibility to scale up our system to generate more logical qubits and construct entangling gates. Section~\ref{discussion} discusses possible experimental realization, the feasibility of our proposal with respect to some experimental parameters, and a subtle comparison between our computation protocols with topological quantum computing (TQC). Finally, we conclude our work in Sec.~\ref{conc}.

\section{Background}

\subsection{Floquet Formalism and Edge Modes}
\label{background1}

Consider a time-periodic (Floquet) Hamiltonian with period $T$, such that $H(t+T)=H(t)$. Since energy is no longer conserved, the spectral properties of the system are instead captured by an analogues quantity called quasienergy \cite{Flo1,Flo2}, defined from the eigenphase of the one-period propagator (Floquet operator) $\mathcal{U}\equiv \mathcal{T} \exp\left(\int_0^T -\frac{\mathrm{i} H(t)}{\hbar} dt\right)$, i.e.,

\begin{equation}
\mathcal{U} |\varepsilon\rangle =\exp\left(-\mathrm{i} \varepsilon T\right) |\varepsilon\rangle \;,
\end{equation}

\noindent where $\mathcal{T}$ is the time-ordering operator, $\varepsilon$ is the quasienergy, and $|\varepsilon\rangle$ is the associated Floquet eigenstate. Since $\varepsilon T$ is only defined up to a modulus of $2\pi$, i.e., $\varepsilon/T$ and $\varepsilon+2\pi n/T$ where $n\in Z$ represent the same solution. As a result, quasienergy is usually defined in $(-\pi/T,\pi/T]$ and forms the so-called Floquet Brillouin zone, which is analogous to quasimomentum Brillouin zone in spatially periodic systems. The periodicity of the quasienergy Brillouin zone is mainly responsible for the existence of edge modes  at quasienergy $\pi/T$ \cite{opt,R1,R2,kk1,kk2,pim,RG} and anomalous edge states \cite{AES,AES2,AES3}. The former is especially relevant to this work, and will thus be elaborated further.

There are two types of edge modes, namely, fermionic and Majorana (half-fermionic) edge modes. In the second quantization language, we define $\Psi_\varepsilon$ as a fermionic mode associated with quasienergy $\varepsilon/T$. Namely, given a reference state $|R\rangle$ satisfying $\mathcal{U}|R\rangle = |R\rangle$, a Floquet eigenstate with quasienergy $\varepsilon/T$  can be constructed as $|\varepsilon\rangle = \Psi_\varepsilon^\dagger |R\rangle$.

In systems possessing chiral symmetry \cite{TSF,TSF3,Derek1} with $\Gamma \mathcal{U} \Gamma^\dagger =\mathcal{U}^\dagger$ for some unitary chiral operator $\Gamma$, quasienergies are guaranteed to come in pairs. That is, associated with a fermionic mode $\Psi_\varepsilon$ at quasienergy $\varepsilon/T$, there exists another fermionic mode $\Psi_{-\varepsilon}=\Psi_\varepsilon \Gamma$ at quasienergy $-\varepsilon/T$. In particular, when $\varepsilon=0$ ($\pi/T$), chiral symmetry dictates that the quasienergy becomes degenerate, i.e., there must exist two fermionic zero ($\pi$) modes $\Psi_0^A$ and $\Psi_0^B$ ($\Psi_\pi^A$ and $\Psi_\pi^B$) related to each other by $\Psi_0^A=\Psi_0^B \Gamma$ ($\Psi_\pi^A=\Psi_\pi^B \Gamma$).

\begin{figure}
	\begin{center}
		\includegraphics[scale=0.45]{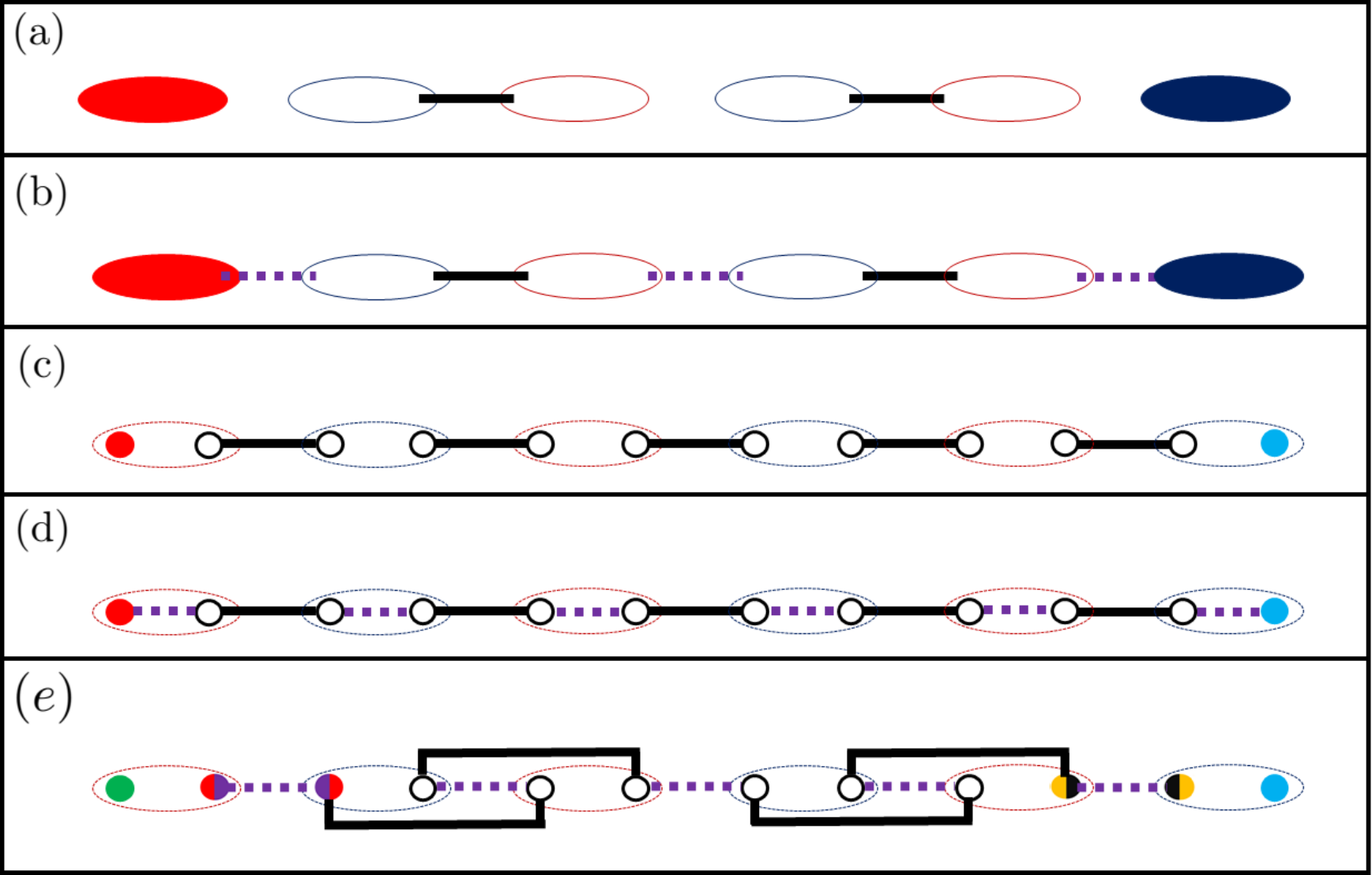}
	\end{center}
	\caption{(color online). Illustration of fermionic zero and $\pi$ modes (coloured ellipses in panel (a) and (b)) and Majorana zero and $\pi$ modes (coloured circles in panel (c) and (d)). Red and blue ellipses represent fermions at sublattice A and B respectively, which can be further broken down into two Majorana fermions (Majoranas), as depicted in circles. {Purple dotted lines and black solid lines denote two different strengths of coupling between two fermions or Majoranas}. Zero modes in panel (a) and (c) arise due to uncoupled fermions or Majoranas, while $\pi$ modes in panel (b) and (d) arise due to the magnitude difference between the purple and black coloured coupling on the fermions or Majoranas near the boundaries. {  Panel (e) illustrates the Hamiltonian Eq.~(\ref{model}) and its associated Majorana zero and $\pi$ modes in the ideal case. Purple dotted lines and black solid lines denote terms originating from $H_1$ and $H_2$ respectively, filled circles mark the Majorana modes, and half-filled circles denote superposition of Majorana modes (see Eq.~(\ref{eq14}) for expressions of edge modes involving superpositions of Majorana fermions at different sublattice sites).}}
	\label{zpi}
\end{figure}

On the other hand, superconducting systems usually also possess an inherent particle-hole symmetry. This associates a fermionic mode $\Psi_\varepsilon$ at quasienergy $\varepsilon/T$ with \emph{the conjugate} of another fermionic mode $\Psi_{-\varepsilon}$ at quasienergy $-\varepsilon/T$, i.e., $\Psi_\varepsilon=\Psi_{-\varepsilon}^\dagger$. As a direct consequence, $\gamma_0\equiv\Psi_0$ and $\gamma_\pi\equiv\Psi_\pi$ become Hermitian, and are thus termed Majorana zero and $\pi$ modes respectively. Since the Floquet operator $\mathcal{U}$ (when expanded) can only contain terms of the form $\Psi^\dagger\Psi$, where $\Psi$ is a complex fermion, Majorana zero ($\pi$) modes should come in pairs as $\gamma_0^{(1)}$ and $\gamma_0^{(2)}$ ($\gamma_\pi^{(1)}$ and $\gamma_\pi^{(2)}$) so as to be able to form a complex fermion $\Psi_0^{(c)}=\gamma_0^{(1)}+\mathrm{i}\gamma_0^{(2)}$ ($\Psi_\pi^{(c)}=\gamma_\pi^{(1)}+\mathrm{i}\gamma_\pi^{(2)}$). In this sense, Majorana zero and $\pi$ modes are clearly half-fermions and fundamentally different from the fermionic zero and $\pi$ modes induced by the less subtle chiral symmetry alone.

Figure~\ref{zpi} depicts zero and $\pi$ edge modes when a gapped system is subject to open boundaries. In particular, fermionic and Majorana modes highlighted above are localized near the systems' left or right boundaries.  By definition, fermionic or Majorana zero modes commute with the Floquet operator $\mathcal{U}$, whereas fermionic or Majorana $\pi$ modes anticommute with the Floquet operator $\mathcal{U}$. Though not pursued in this work, we note that the $\pi$ edge modes being anticommuting with $\mathcal{U}$ offers a dynamical-decoupling scenario from within the system dynamics itself and they are thus expected to be even more robust than zero edge modes against certain noise.
It should be also noted that while fermionic and Majorana zero modes can also emerge in static systems by the same mechanism elucidated above, fermionic and Majorana $\pi$ modes can only exist in Floquet systems due to the periodicity of quasienergy.

\subsection{Floquet adiabatic process and holonomy}
\label{holonomy}

Let $H(t,\lambda)$ be time-periodic with period $T$ and depending also on a tunable parameter $\lambda$. If Floquet eigenstates are not degenerate, then a Floquet adiabatic process is accomplished by slowly tuning $\lambda$ from a certain initial value $\lambda_0$ at time zero to a final value $\lambda_\tau$ at time $\tau=M T$, such that a state initially prepared in a Floquet eigenstate with quasienergy $\epsilon_n(\lambda_0)$ will evolve with $\lambda$ as an {  instantaneous} Floquet eigenstate with quasienergy $\epsilon_n(\lambda)$ \cite{Hailong}.  It is convenient to assume that $\lambda$ is only tuned stroboscopically at the beginning of each new driving period, such that $\lambda\equiv\lambda(s)$ when $sT\leq t<(s+1)T$. Adiabaticity then requires $\tau/T=M\gg 1$ as well as other conditions involving the gap of the Floquet states versus $\hbar/T$ \cite{Hailong}.

Floquet adiabatic holonomy arises from a Floquet adiabatic process in which $H(t,\lambda_\tau)=H(t,\lambda_0)$ and its associated Floquet operator $\mathcal{U}(\lambda)$ always possesses degenerate Floquet states throughout the adiabatic cycle. For each quasienergy $\varepsilon_n$, we can thus define a column vector containing all of its degenerate Floquet eigenstates as $|\varepsilon_n\rangle \equiv\left(|\varepsilon_{n,1}\rangle ,\cdots , |\varepsilon_{n,k_n}\rangle\right)^T$, where $k_n$ is the number of degeneracy associated with $\varepsilon_n$. As detailed in Appendix~\ref{app1}, the evolution of a Floquet eigenstate $|\varepsilon_n \rangle$ of $\mathcal{U}(\lambda_0)$ after one adiabatic cycle is given by

\begin{equation}
|\varepsilon_n (\lambda_\tau)\rangle =\mathcal{P} \exp\left(-\mathrm{i}\oint \left[\mathcal{A}_n+\Omega_n+\varepsilon_n T\right] d\lambda \right) |\varepsilon_n (\lambda_0)\rangle\;,
\label{hol}
\end{equation}

\n where $\mathcal{P}$ is the path ordering operator, $\mathcal{A}_n$ and $\Omega_n$ are defined in Appendix~\ref{app1}, and the closed integration is used since the Hamiltonian returns to itself after one adiabatic cycle.

The first term in the exponential of Eq.~(\ref{hol}) is the non-Abelian Berry matrix, while the second term represents the explicit monodromy \cite{geo1,geo2}, i.e., permutation/braiding in the degenerate subspace, induced by the holonomy. The summation of the first two terms gives rise to the total non-Abelian geometric phase of the system, whereas the last term denotes the dynamical phase contribution. In particular, since the geometric phase appears as a matrix, Eq.~(\ref{hol}) may in general induce a nontrivial rotation of $|\varepsilon_n(\lambda_0)\rangle$ within the degenerate subspace, so that $|\varepsilon_n(\lambda_0)\rangle$ and $|\varepsilon_n(\lambda_\tau)\rangle$ are not simply related by an overall phase as in the nondegenerate (Abelian) case. This is the basic idea behind holonomic and topological quantum computation (HQC and TQC), which we have now extended to Floquet systems. Equation~(\ref{hol}) also makes it clear why holonomic quantum computation with topologically zero modes and $\pi$ modes are of special interest: the dynamical phase contribution can be clearly separated out if the zero or $\pi$ modes persist throughout the adiabatic process. That is, the dynamical phase $ -\int_{\lambda_0}^
{\lambda_\tau} \varepsilon_n T d\lambda  = M \varepsilon_n T$ is equivalent to zero given that $\varepsilon_n=0$ or $\pi/T$
and that $M$ is even (that is, if the adiabatic process takes even multiples of driving periods).

\section{Description of the model}
\label{models}

The general model we will be using throughout this work describes a 1D time-periodic $p$-wave superconducting superlattice with alternating real and imaginary hopping as well as pairing at every half period, i.e.,

\begin{eqnarray}
H(t) &=& \begin{cases}
H_1 & \text{ for } (m-1)T < t\leq (m-1/2)T \\
H_2 & \text{ for } (m-1/2)T < t\leq mT
\end{cases} \;, \label{model} \\
H_1 &=& \sum_{i} \left(-J_{\mathrm{intra},i} c_{B,i}^\dagger c_{A,i} -J_{\mathrm{inter},i} c_{A,i+1}^\dagger c_{B,i}\right. \nonumber \\
&& \left. +\Delta_{\mathrm{intra},i} c_{B,i}^\dagger c_{A,i}^\dagger +\Delta_{\mathrm{inter},i} c_{A,i+1}^\dagger c_{B,i}^\dagger + h.c. \right) \;, \nonumber \\
H_2 &=& \sum_{i} \left(-\mathrm{i} j_{\mathrm{intra},i} c_{B,i}^\dagger c_{A,i}-\mathrm{i} j_{\mathrm{inter},i} c_{A,i+1}^\dagger c_{B,i}\right. \nonumber \\
&& \left. +\mathrm{i} \delta_{\mathrm{intra},i} c_{B,i}^\dagger c_{A,i}^\dagger +\mathrm{i} \delta_{\mathrm{inter},i} c_{A,i+1}^\dagger c_{B,i}^\dagger + h.c. \right) \;, \nonumber \\
\label{H}
\end{eqnarray}

\noindent where $c_{A,i}$ ($c_{B,i}$) and $c_{A,i}^\dagger$ ($c_{B,i}^\dagger$) denote the fermion creation and annihilation operators at sublattice A (B) of lattice site $i$ respectively, $J_{\mathrm{intra},i}$, $J_{\mathrm{inter},i}$, $j_{\mathrm{intra},i}$, and $j_{\mathrm{inter},i}$ denote intra- and inter-lattice hopping strength at site $i$ at different half of the period, $\Delta_{\mathrm{intra},i}$, $\Delta_{\mathrm{inter},i}$, $\delta_{\mathrm{intra},i}$, and $\delta_{\mathrm{inter},i}$ are the intra- and inter-lattice pairing strength at site $i$ at different half of the period. The total number of lattice sites is denoted as $N$, which is finite in our actual calculations under open boundary conditions.  By construction, $T$ is the time period of the above periodically-quenched  Hamiltonian. Unless otherwise specified, we take $J_{\mathrm{intra},i}=J_1$, $J_{\mathrm{inter},i}=J_2$, $j_{\mathrm{intra},i}=j_1$, $j_{\mathrm{inter},i}=j_2$, $\Delta_{\mathrm{intra},i}=\Delta_1$, $\Delta_{\mathrm{inter},i}=\Delta_2$, $\delta_{\mathrm{intra},i}=\delta_1$, and $\delta_{\mathrm{inter},i}=\delta_2$ for all $i=1,\cdots,N$.
Each of $H_1$ or $H_2$ itself depicted in Eq.~(\ref{H}) represents a static dimerized Kitaev chain.  In the absence of sublattice degree of freedom, i.e., by taking $J_1=J_2$, $\Delta_1=\Delta_2$, $j_1=j_2$, and $\delta_1=\delta_2$, Eq.~(\ref{model}) reduces to a time-periodic Kitaev Hamiltonian, which is known to possess Majorana zero  edge modes \cite{Kit} under suitable parameter values. Due to the sublattice degree of freedom, the SSH-like zero (quasi) energy edge modes \cite{SSH} are also expected.

In general, Kitaev- and SSH-like zero edge modes will compete with each other, and only one of them can exist for a given set of system parameters. This competition can be well understood in terms of an integer topological invariant \cite{SSHK1,SSHK2,kk2}. On the other hand, since our system is periodically quenched, Kitaev- or SSH-like edge modes at quasienergy $\pi/T$ may also exist, which are governed by a separate integer topological invariant \cite{FTC2}. As a result, while only one type of edge modes can emerge at quasi-energy zero or $\pi/T$, it is possible to find certain parameter windows for which two different types of edge modes coexist, one at quasienergy zero, while the other at quasienergy $\pi/T$ \cite{kk2,kk1}.

\subsection{Symmetry analysis}
\label{sym}

To gain more insights into our working model, we first rewrite Eq.~(\ref{H}) in the Nambu-momentum representation as follows:

\begin{eqnarray}
H_l &=& \sum_{k>0} \Psi_k^\dagger h_{l,k} \Psi_k \;, \nonumber \\
h_{1,k} &=& - \tau_z  J(k)\cdot \sigma  + \tau_y \Delta(k)\cdot \sigma  \;, \nonumber \\
h_{2,k} &=& -j(k)\cdot \sigma +\tau_x \delta(k) \cdot \sigma  \;.
\label{hmom}
\end{eqnarray}

\noindent where $\Psi_k^\dagger=\left(c_{A,k}^\dagger,c_{B,k}^\dagger,c_{A,-k},c_{B,-k}\right)$, $l\in\left\lbrace1,2\right\rbrace$, $\sigma_i$ and $\tau_i$ are Pauli matrices in the sublattice and particle-hole degrees of freedom respectively. Other terms used above are given by
  \begin{eqnarray}
  J(k)\cdot \sigma & =& \left(J_1 +J_2 \cos k\right) \sigma_x -J_2 \sin k \sigma_y\;, \nonumber \\
  j(k)\cdot \sigma& =& \left(j_1 -j_2 \cos k \right) \sigma_y -j_2 \sin k \sigma_x\;, \nonumber \\
  \Delta(k)\cdot \sigma& =& \left(\Delta_1 -\Delta_2 \cos k \right)\sigma_y -\Delta_2 \sin k \sigma_x\;,\nonumber \\
   \delta(k)\cdot \sigma & =&\left(\delta_1 -\delta_2 \cos k\right) \sigma_y -\delta_2 \sin k \sigma_x \;.
    \end{eqnarray}
   To analyze the symmetry, it is convenient to consider the momentum space Floquet operator in a symmetric time frame \cite{TSF,TSF2,TSF3,Derek1} as

\begin{eqnarray}
U_k &=& \hat{F}_k \hat{G}_k\;, \label{flo}\\
\hat{F}_k &=& \exp\left(-\mathrm{i}h_{1,k} T/4 \right)\times \exp\left(-\mathrm{i}h_{2,k} T/4 \right) \;, \nonumber \\
\hat{G}_k &=& \exp\left(-\mathrm{i}h_{2,k} T/4 \right)\times \exp\left(-\mathrm{i}h_{1,k} T/4 \right) \;.
\end{eqnarray}

It can be checked that Eq.~(\ref{flo}) possesses sublattice chiral symmetry since $\Gamma \hat{F}_k \Gamma^\dagger = \hat{G}_k^\dagger$ with $\Gamma=\sigma_z$ \cite{TSF,Derek1}. As expected for a typical superconducting system, Eq.~(\ref{flo}) also possesses particle-hole symmetry given by $\mathcal{P} U_k \mathcal{P}^{-1}=U_{-k}$, where $\mathcal{P}=\tau_x \mathcal{K}$ and $\mathcal{K}$ is the complex conjugation operator \cite{FTC2,kk2}. The presence of both chiral and particle-hole symmetries also implies the existence of time reversal symmetry dictated by the operator $\mathcal{T}=\sigma_z \tau_x \mathcal{K}$, which is easily verified in the symmetric time frame according to $\mathcal{T} h(k,t) \mathcal{T}^{-1}=h(-k,T-t)$, where $h(k,t)$ is the {full} time-dependent Hamiltonian depicted by Eq.~(\ref{model}) in the momentum space \cite{FTC2,kk2}. Our working system thus belongs to the BDI class according to the Altland-Zirnbauer classification scheme \cite{AZ}, which is characterized by a $Z\times Z$ topological invariant \cite{FTC2}.

\subsection{$Z\times Z$ topological invariant}
\label{topi}

As a result of the chiral symmetry, we can identify the $Z\times Z$ topological invariants by combining some techniques from Ref.~\cite{TSF,TSF3,SSHK1}. First, we change Eq.~(\ref{hmom}) to a canonical basis \cite{SSHK1} by applying a unitary transformation with

\begin{equation}
U=\frac{\left(1+\sigma_x\right)+\tau_z\left(1-\sigma_x\right)}{2} \times \frac{\left(1+\tau_x\right)+\sigma_z\left(1-\tau_x\right)}{2} \;,
\end{equation}

\n so that $U^\dagger \Gamma U=\tau_z$. Next, we follow Ref.~\cite{TSF3} and write $\hat{F}_k$ in this basis as a block matrix, i.e.,

\begin{equation}
\hat{F}_k\hat{=} \left(\begin{array}{cc}
a(k) & b(k) \\
c(k) & d(k)
\end{array} \right) \;,
\end{equation}

\n where each block is a $2\times 2$ matrix.

\begin{figure*}
	\begin{center}
		\includegraphics[scale=0.5]{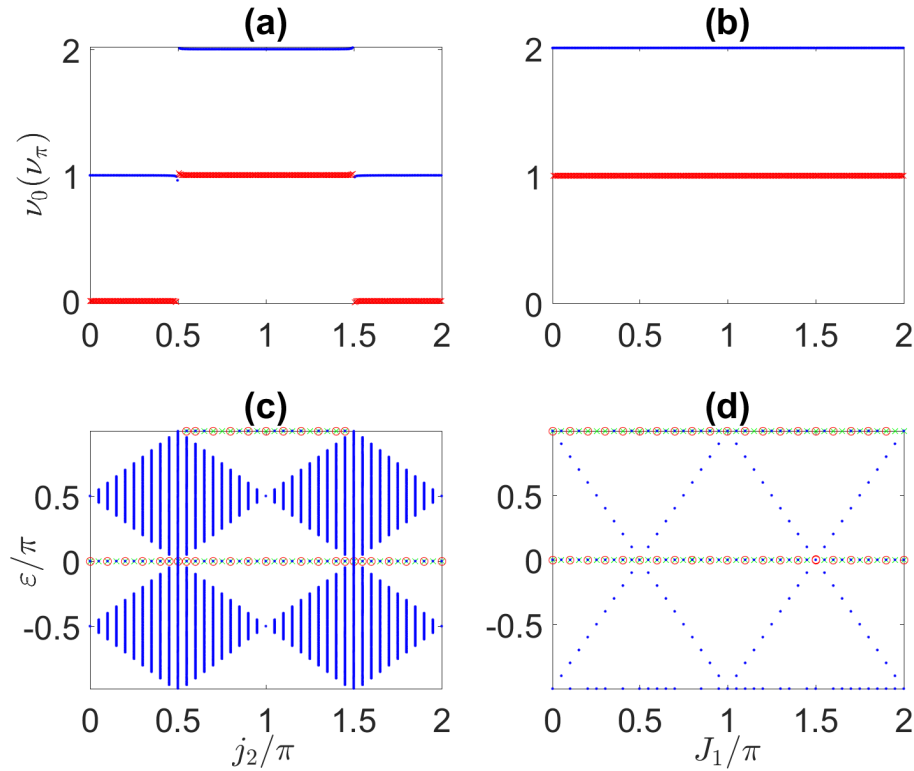}
	\end{center}
	\caption{(color online). (a) and (b): Topological invariants $\nu_0$ (blue) and $\nu_\pi$ (red) versus the system parameters, predicting totally three pairs of Majorana edge states, with two of them forming a pair of SSH-like edge modes.  (c) and (d): The associated Floquet spectrum under OBC, where the left and right localized edge states are marked with green crosses and red circles respectively. The system parameters are chosen as $j_1=\delta_1=\delta_2=0$, $J_1=\Delta_1=J_2=\Delta_2=\pi/4$ (panel (a) and (c)), $j_2=\pi$ (panel (b) and (d)).}
	\label{ZZ}
\end{figure*}

The number of edge states at quasienergy zero and $\pi$ can then respectively be found by calculating the topological invariants \cite{TSF3}

\begin{eqnarray}
\nu_0 &=& \frac{1}{2\pi \mathrm{i}} \int_{-\pi}^\pi dk \mathrm{Tr} \left(b^{-1}\frac{d}{dk} b\right) \;, \nonumber \\
\nu_\pi &=& \frac{1}{2\pi \mathrm{i}} \int_{-\pi}^\pi dk \mathrm{Tr} \left(d^{-1}\frac{d}{dk} d\right) \;. \label{topin}
\end{eqnarray}

\n The topological invariants $\nu_0$ and $\nu_\pi$ under some representative parameter values are depicted in Fig.~\ref{ZZ}, along with their associated Floquet eigen-spectrum under open boundary conditions (OBC) { \cite{note2}}. There, $\nu_0=1$ ($\nu_\pi=1$) is associated with the presence of Kitaev-like edge states at quasienergy zero ($\pi/T$), which predicts only one Majorana zero ($\pi$) edge mode at each edge. On the other hand, SSH-like edge states, being complex-fermionic in nature, can be broken down into two Majorana zero ($\pi$) modes and thus emerge whenever $\nu_0=2$ ($\nu_\pi=2$). The parameter window for which $\nu_0=2$ and $\nu_\pi=1$ will be used in this work, because the coexistence of Kitaev- and SSH-like edge states will prove to be essential for our encoding and manipulation of the logical qubits we obtain.

\section{Edge-modes based qubit encoding}
\label{encoding}

\subsection{Edge modes in the Majorana representation}

In Sec.~\ref{topi}, we have shown that for certain parameter windows, two different species of edge modes, originating from sublattice and particle-hole symmetry-protected topology respectively, may coexist on one single quantum wire. To elucidate on the application of these edge modes for quantum computation, it is convenient to first define (Hermitian) Majorana operators as follows,

\begin{eqnarray}
\gamma_{s,i}^\alpha &=& c_{s,i}+c_{s,i}^\dagger \;,\nonumber \\
\gamma_{s,i}^\beta &=& \mathrm{i} \left(c_{s,i}-c_{s,i}^\dagger \right) \;,
\end{eqnarray}

\noindent where $s\in\left\lbrace A, B\right\rbrace$ and $i=1,\cdots , N$. Moreover, to simplify our analysis, we will take the following parameter values: $j_1=\delta_1=\delta_2=0$, $J_1 T=J_2 T=\Delta_1 T=\Delta_2 T=\pi/2$, and $j_2 T=2\pi$, which from here onwards shall be referred to as \emph{the ideal case}. It should be stressed however that such fine tuning is not necessary in the actual implementation, and the results we present in the following still hold under small deviations from the ideal case.

In the ideal case, Eq.~(\ref{H}) can be written in terms of Majorana operators as

\begin{eqnarray}
H_1 T &=& \sum_{i} \mathrm{i} \frac{\pi}{2}\left( \gamma_{B,i}^\alpha \gamma_{A,i}^\beta+ \gamma_{A,i+1}^\alpha \gamma_{B,i}^\beta\right)  \;, \nonumber \\
H_2 T &=& -\sum_i \mathrm{i} \pi \left(\gamma_{A,i+1}^\alpha \gamma_{B,i}^\alpha + \gamma_{A,i+1}^\beta \gamma_{B,i}^\beta\right) \;,
\label{Hm}
\end{eqnarray}

\noindent which is graphically represented in Fig.~\ref{zpi}(e). We can then find exact expressions for a pair of Majorana $\pi$ modes and two pairs of Majorana zero modes given by

\begin{eqnarray}
\gamma_{\pi}^L &=& \frac{1}{\sqrt{2}}\left(\gamma_{A,1}^\beta+\gamma_{B,1}^\alpha\right) \;, \nonumber \\
\gamma_\pi^R &=& \frac{1}{\sqrt{2}}\left(\gamma_{A,N}^\beta-\gamma_{B,N}^\alpha\right) \;, \nonumber \\
\gamma_{0,1}^L &=& \gamma_{A,1}^\alpha \;, \nonumber \\
\gamma_{0,1}^R &=& \frac{1}{\sqrt{2}}\left(\gamma_{A,N}^\beta+\gamma_{B,N}^\alpha\right) \;, \nonumber \\
\gamma_{0,2}^L &=& \frac{1}{\sqrt{2}}\left(\gamma_{A,1}^\beta-\gamma_{B,1}^\alpha\right) \;, \nonumber \\
\gamma_{0,2}^R &=& \gamma_{B,N}^\beta \;,
\label{eq14}
\end{eqnarray}

\noindent which satisfy
\begin{eqnarray}
\mathcal{U}^\dagger \gamma_\pi^s \mathcal{U}& =& -\gamma_\pi^s \;, \nonumber\\
 \mathcal{U}^\dagger \gamma_{0,l}^s \mathcal{U} &=&\gamma_{0,l}^s \;,
  \end{eqnarray}
 where $s\in\left\lbrace L,R \right\rbrace$, $l\in\left\lbrace 1,2\right\rbrace$, and $\mathcal{U}=\exp(-\mathrm{i} H_2 T/2)\times \exp(-\mathrm{i} H_1 T/2)$ is the Floquet operator.  Take the edge modes  localized at the left end as examples.
 The $\pi$ mode $\gamma_{\pi}^L$ is a superposition of two Majorana operators involving both $A$ and $B$ sublattices and is exclusively localized at the very first lattice. The two zero modes $\gamma_{0,1}^L$ and  $\gamma_{0,2}^L$ are also localized at the first site, one involving the real part of $c_{A,1}$ only and the other as a different superposition of the two Majorana operators involving both $A$ and $B$ sublattices.

  The pair of Majorana $\pi$ modes at opposite ends can be fused to form a nonlocal fermion $f_\pi =\gamma_\pi^L +\mathrm{i} \gamma_\pi^R$. On the other hand, at each end of the wire, there are two Majorana zero modes as shown above. They can locally form a fermion (hence the SSH-like zero edge modes), which can be denoted by $f_0^L=\gamma_{0,1}^L +\mathrm{i} \gamma_{0,2}^L$ at the left edge (or $f_0^R=\gamma_{0,1}^R +\mathrm{i} \gamma_{0,2}^R$ at the right edge) {  \cite{note}}.  Our encoding scheme elaborated below shall use both the nonlocal fermion and the two local fermions.

\subsection{Qubit encoding}
\label{encode}

Given that the system should preserve the total fermion parity, the three edge fermions defined above can be exploited to encode up to two logical qubits. Let $|n_0^L n_\pi n_0^R\rangle$ be a simultaneous eigenstate of three fermion parity operators $\mathcal{P}_\pi=\mathrm{i} \gamma_\pi^L \gamma_\pi^R$, $\mathcal{P}_0^L=\mathrm{i} \gamma_{0,1}^L \gamma_{0,2}^L$, and $\mathcal{P}_0^R=\mathrm{i} \gamma_{0,1}^R \gamma_{0,2}^R$. Then, by definition we have

\begin{eqnarray}
\mathcal{P}_\pi |n_0^L n_\pi n_0^R\rangle &=& (-1)^{n_\pi}|n_0^L n_\pi n_0^R\rangle \;, \nonumber \\
\mathcal{P}_0^L |n_0^L n_\pi n_0^R\rangle &=& (-1)^{n_0^L}|n_0^L n_\pi n_0^R\rangle \;, \nonumber \\
\mathcal{P}_0^R |n_0^L n_\pi n_0^R\rangle &=& (-1)^{n_0^R}|n_0^L n_\pi n_0^R\rangle \;,
\end{eqnarray}

\noindent where $n_\pi,n_0^L,n_0^R\in \left\lbrace 0,1 \right\rbrace$. There are now $8$ simultaneous eigenstates of $\mathcal{P}_\pi$, $\mathcal{P}_0^L$, and $\mathcal{P}_0^R$.  The total parity conservation divides this eight-dimensional Hilbert space into two four-dimensional parity preserving subspaces. The odd and even parity subspace are respectively spanned by $\left\lbrace|0 0 1 \rangle, |0 1 0\rangle, |1 0 0\rangle, |1 1 1\rangle \right\rbrace$ and $\left\lbrace|0 0 0 \rangle, |0 1 1\rangle, |1 1 0\rangle, |1 0 1\rangle \right\rbrace$. Without loss of generality, in this work we assume that the system is initialized in the even parity subspace. This allows us to define the four qubit basis states with $|00\rangle \equiv |000\rangle$, $|01\rangle \equiv |0 1 1\rangle$, $|10\rangle \equiv |1 1 0 \rangle$, and $|11\rangle \equiv | 10 1\rangle$.  These four qubit states, which represent the basis states of two logical qubits,  are related to each other by

\begin{eqnarray}
|01\rangle &=& \gamma_\pi^L \gamma_{0,1}^R |0 0\rangle \;, \nonumber \\
|10\rangle &=& \gamma_{0,2}^L \gamma_\pi^L |0 0\rangle \;, \nonumber \\
|11\rangle &=& \gamma_{0,2}^L \gamma_{0,1}^R |0 0\rangle \;.
\label{qubits}
\end{eqnarray}

%\textit{One-qubit encoding.---}By further imposing another constraint that either the parity of $\mathcal{P}_0^L$ or $\mathcal{P}_0^R$ is fixed, the dimension of the computational Hilbert is further reduced into two. For instance, in the even parity subspace of $\mathcal{P}_0^L$, we may define the two qubit states $|0\rangle \equiv |000\rangle$ and $|1\rangle \equiv |011\rangle$. In this case, $\mathcal{P}_0^L$ serves as an additional stabilizer operator which may be exploited to keep track of and correct errors arising from some chiral symmetry breaking perturbations.

\section{Holonomic quantum computation with edge modes}
\label{compute}

Having shown how logical qubits can be encoded in our system, we now investigate which logical gate operations can be implemented. For Majorana-based qubits, topologically protected gate operations can be carried out through braiding between a pair of Majorana modes \cite{tqc2}. Assuming that all pairs of Majorana modes in a given system can be braided, all Clifford, i.e., Hadamard, CNOT, and phase, gates can in principle be implemented \cite{tqc2,Clif1,Clif2}. However, in many proposed systems hosting Majorana modes, especially those in 1D setups, braiding some pairs of Majorana modes may be challenging, {especially if they are separated too far apart. For example, in 1D systems such as those studied in Ref.~\cite{RG, wire}, a single qubit requires two pairs of Majorana modes located at two opposite edges. As such, braiding one Majorana mode from one edge with that from the other edge may be quite difficult to carry out in practice, which in turn hinders the realization of universal quantum computation.}

Recognizing that relying exclusively on nonlocal Majorana qubits is still a big challenge for quantum computation purposes, our qubit encoding scheme outlined in the previous subsection represents a hybrid scenario with both local and nonlocal fermions. The advantage of involving local fermions in our encoding is that it allows more pairs of Majorana modes to be easily braided. As seen below, this feature leads to the implementation of a larger set of gate operations, at least in principle. In the following, we explicitly present the protocols to implement some gate operations by braiding between different pairs of Majorana modes.  This is done by adiabatically deforming the system's Hamiltonian in closed cycles, in the spirit of holonomic quantum computation \cite{HQC,TQCBook}.

\subsection{Phase gate and Pauli $Z$ gate}
\label{prot1}

 With the two-qubit encoding introduced in Sec.~\ref{encode}, single phase gate and Pauli $Z$ gate (up to a global phase factor) on the first or second qubit individually can be obtained by braiding $\gamma_{0,1}^L$ and $\gamma_{0,2}^L$ or braiding $\gamma_{0,1}^R$ and $\gamma_{0,2}^R$ once and twice respectively.  In terms of {  braiding unitaries $U_s$, the phase and Pauli $Z$ gate are respectively} $P_s\equiv U_s=\exp\left[(\pi/4) \gamma_{0,2}^s \gamma_{0,1}^s\right]$ and $Z_s \equiv U_s^2=\exp\left[(\pi/2) \gamma_{0,2}^s \gamma_{0,1}^s\right]$, where $s=L$ ($s=R$). This can be verified by applying $P_s$ and $Z_s$ directly to Eq.~(\ref{qubits}), with the obvious identity

\begin{equation}
\exp\left(\theta \gamma_{0,1}^s \gamma_{0,2}^s \right)= \cos\theta +\sin\theta \gamma_{0,1}^s \gamma_{0,2}^s \;.
\label{id}
\end{equation}
\n Indeed, identity Eq.~(\ref{id}) {  can also be employed to see that $U_s$ satisfies the usual relation $U_s^\dagger \gamma_{0,1}^s U_s=-\gamma_{0,2}^s$ and $U_s^\dagger \gamma_{0,2}^s U_s=\gamma_{0,1}^s$}.

\begin{figure*}

%\ffigbox[\FBwidth]
	\begin{center}
		\includegraphics[scale=0.8]{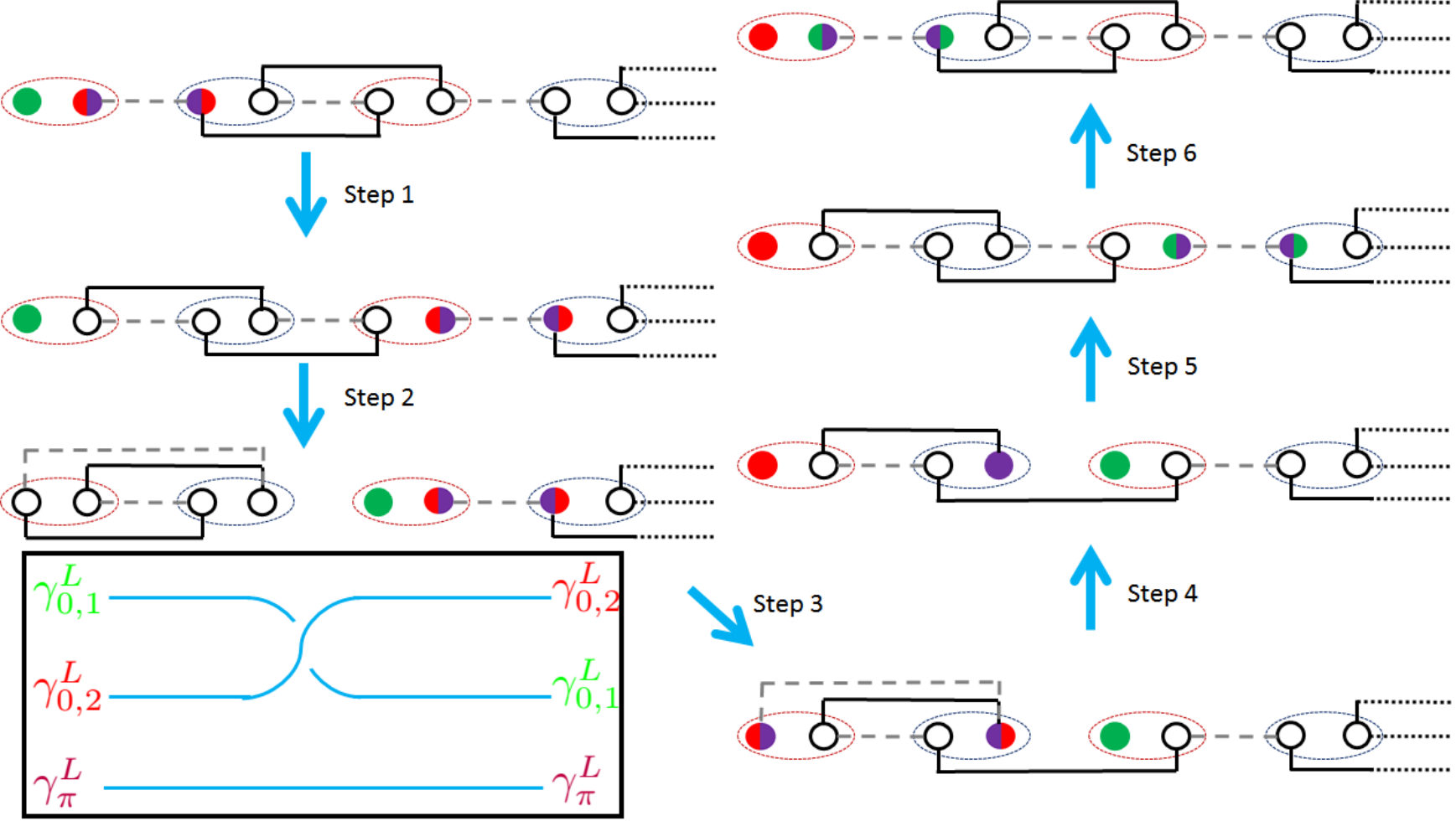}
	\end{center}
	\caption{(color online). Schematic of the holonomic protocol to braid $\gamma_{0,1}^L$ and $\gamma_{0,2}^L$. Only the first two lattice sites are shown. Red and blue ellipses represent sublattice A and B respectively, with two circles at each ellipse are the associated Majorana operators. Coloured circles denote the Majorana modes as described in the inset. Some Majorana modes are superposition of two Majorana operators, which are represented by half-coloured circles. Black solid and gray dashed  lines denote the coupling between two Majorana operators due to $H_2$ and $H_1$ respectively.}
	\label{braidp1}
\end{figure*}

We now present below the details of our protocol to realize the braiding unitary $U_s$. { Starting with the Hamiltonian of Eq.~(\ref{model}), each step below amounts to varying the coupling between pairs of lattice sites, whose effect in Majorana representation is illustrated in Fig.~\ref{braidp1}}. In order to simplify our discussion, we focus on the ideal case, which allows us to keep track of the analytical solutions at the end of each step. As will be shown in our numerics later on, however, the result of our protocol still holds even if we tune the system parameters away from the ideal case.  Furthermore, we will only present the protocol to braid $\gamma_{0,1}^L$ and $\gamma_{0,2}^L$ (to find $U_L$). Braiding $\gamma_{0,1}^R$ and $\gamma_{0,2}^R$ can be accomplished in the same fashion, by applying our considerations to the right edge instead. For each step elaborated below, the adiabatic parameter $\phi$ is slowly increased at the beginning of each driving period, starting from $0$ and ending at $\pi/2$ after a total of even number of driving periods.  We only briefly elucidate the output of each step, thus leaving more technical details in Appendix~\ref{app2}.

%\begin{widetext}
%\lipsum[2]
\ \ \
%\end{widetext}

\textit{Step 1.---}{  With $H_1$ untouched, we start the procedure by varying $j_{\mathrm{inter},1} T= \pi(1+\cos\phi)$, $\delta_{\mathrm{inter},1}=-\pi(1-\cos\phi)$, $j_{\mathrm{intra},1} T = \delta_{\mathrm{intra},1} T= \pi\sin\phi$, where $\phi$ is the adiabatic parameter (also in all other steps below). By writing down the resulting Hamiltonian in terms of Majorana operators, as presented in Appendix~\ref{app2}, this step is shown to adiabatically move $\gamma_{0,2}^L$ and $\gamma_\pi^L$ to the second  lattice site, i.e., $\frac{1}{\sqrt{2}}\left(\gamma_{A,1}^\beta\pm\gamma_{B,1}^\alpha\right)$ to $\frac{1}{\sqrt{2}}\left(\gamma_{A,2}^\beta\pm\gamma_{B,2}^\alpha\right)$.}%Adiabatic deformation of $H_1$ and $H_2$ is introduced by setting $\left(j_{\mathrm{inter},1} +\delta_{\mathrm{inter},1}\right) T= 2\pi$,  $\left(J_{\mathrm{intra},1} +\Delta_{\mathrm{intra},1}\right) T= \pi$  and slowly tuning $\left(j_{\mathrm{inter},1} -\delta_{\mathrm{inter},1}\right) T= 2\pi\cos\phi$, $j_{\mathrm{intra},1} T = -\delta_{\mathrm{intra},1} T= \pi\sin\phi$, $J_{\mathrm{inter},1} T=\Delta_{\mathrm{inter},1} T= \frac{\pi}{2}\cos\phi$, and $\left(J_{\mathrm{intra},1} -\Delta_{\mathrm{intra},1}\right) T= -\pi\sin\phi$ with $\phi$ always being the adiabatic parameter in all the steps. The net outcome of this step is to move $\gamma_{0,1}^L$ to the second lattice site, i.e., changing $\gamma_{A,1}^L$ to $\gamma_{A,2}^L$.

\textit{Step 2.---}{  Next, we vary (via $\phi$) $j_{\mathrm{inter},1} T=-\delta_{\mathrm{inter},1} T=\pi \cos\phi$, $j_{\mathrm{intra},1} T= \pi(1+\sin\phi)$, $\delta_{\mathrm{intra},1} T= \pi(1-\sin\phi)$, $J_{\mathrm{intra},1} T= \frac{\pi}{2}(1-\sin\phi)$, $\Delta_{\mathrm{intra},1} T= \frac{\pi}{2}(1+\sin\phi)$, and $J_{\mathrm{inter},1} T=\Delta_{\mathrm{inter},1} T= \frac{\pi}{2}\cos\phi$.  This step results in moving $\gamma_{0,1}^L$ to the second lattice site, i.e., changing $\gamma_{A,1}^L$ to $\gamma_{A,2}^L$}.

\textit{Step 3.---}{  We continue by varying $j_{\mathrm{intra},1} T= \pi(1+\cos\phi)$, $\delta_{\mathrm{intra},1} T= \pi(1-\cos\phi)$, $j_{\mathrm{inter},1}T=-\delta_{\mathrm{inter},1} T=-\mathrm{i} \pi \sin\phi$. At the end of this step, $\gamma_\pi^L=\frac{1}{\sqrt{2}}\left(\gamma_{A,1}^\alpha+\gamma_{B,1}^\beta\right)$ and $\gamma_{0,2}^L=\frac{1}{\sqrt{2}}\left(\gamma_{A,1}^\alpha-\gamma_{B,1}^\beta\right)$. That is, $\gamma_\pi^L$ and $\gamma_{0,2}^L$ return to the first lattice site, but they have transformed to different superpositions of Majorana operators.}

\textit{Step 4.---}{  This step amounts to separating $\gamma_{0,2}^L$ from $\gamma_\pi^L$, which is accomplished by tuning $J_{\mathrm{intra},1} T= \frac{\pi}{2}(1-\cos\phi)$ and $\Delta_{\mathrm{intra},1} T= \frac{\pi}{2}(1+\cos\phi)$, such that $\gamma_\pi^L=\gamma_{B,1}^\beta$ and $\gamma_{0,2}^L=\gamma_{A,1}^\alpha$ at the end of this step.}

\textit{Step 5.---}In this step, $\gamma_{0,1}^L$ and $\gamma_\pi^L$ are turned into superpositions of two Majorana operators. This is done by tuning $j_{\mathrm{inter},1}T=-\delta_{\mathrm{inter},1}T=-\pi \exp\left[\mathrm{i} (\pi/2+\phi)\right]$ and $J_{\mathrm{inter},1}T=\Delta_{\mathrm{inter},1}T=\frac{\pi}{2}\sin\phi$, which leads to $\gamma_\pi^L=\frac{1}{\sqrt{2}}\left(\gamma_{A,2}^\beta+\gamma_{B,2}^\alpha\right)$ and $\gamma_{0,1}^L=-\frac{1}{\sqrt{2}}\left(\gamma_{A,2}^\beta-\gamma_{B,2}^\alpha\right)$ at the end of the step.

\textit{Step 6.---}Finally, $H_1$ and $H_2$ are returned to their original forms. This is done by tuning $j_{\mathrm{intra},1}T=\delta_{\mathrm{intra},1}T=\pi\cos\phi$, {  $j_{\mathrm{inter},1} T= \pi(1+\sin\phi)$, and $\delta_{\mathrm{inter},1} T= -\pi(1-\sin\phi)$}, which results in $\gamma_\pi^L=\frac{1}{\sqrt{2}}\left(\gamma_{A,1}^\beta+\gamma_{B,1}^\alpha\right)$ and $\gamma_{0,1}^L=-\frac{1}{\sqrt{2}}\left(\gamma_{A,1}^\beta-\gamma_{B,1}^\alpha\right)$ at the end of the step.

\begin{figure*}
	\begin{center}
		\includegraphics[scale=0.5]{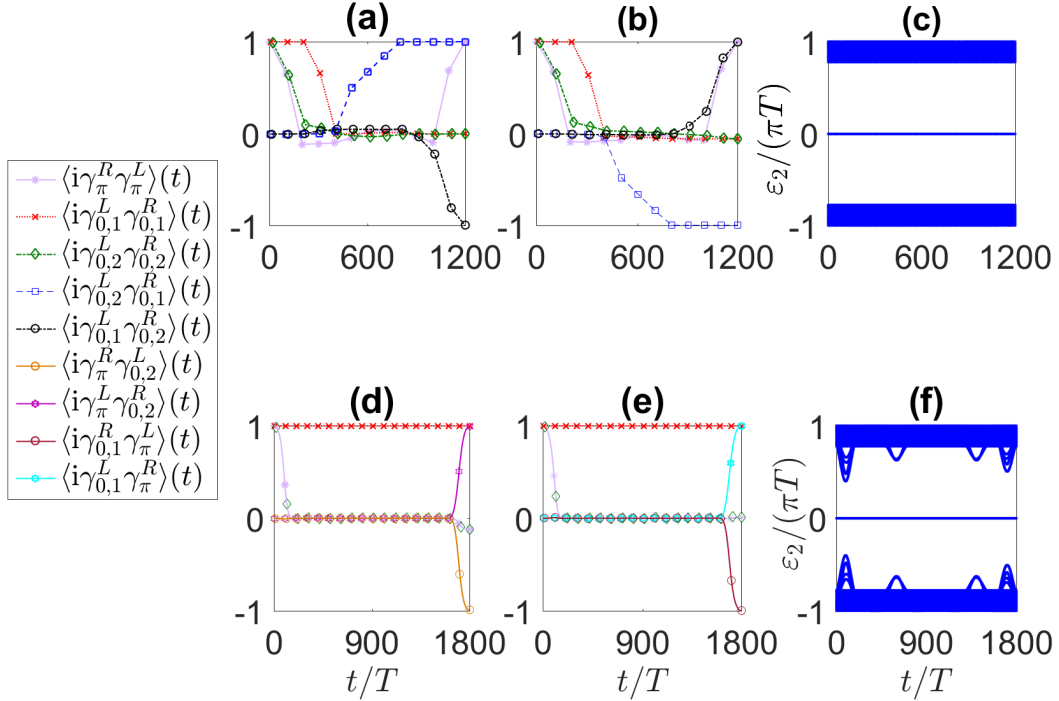}
	\end{center}
	\caption{(color online). Evolution of the Majorana correlation functions under the implementation of the protocol described in Sec.~\ref{prot1} and Sec.~\ref{prot2} to braid (a) $\gamma_{0,1}^L$ and $\gamma_{0,2}^L$, (b) $\gamma_{0,1}^R$ and $\gamma_{0,2}^R$, (c) $\gamma_{0,2}^L$ and $\gamma_\pi^L$, or (d) $\gamma_{0,1}^R$ and $\gamma_\pi^R$. (c) and (f) show the evolution of {  the instantaneous eigenphase spectrum associated with two-period Floquet operator $\mathcal{U}^2$} under the adiabatic parameter tuning described in Sec.~\ref{prot1} and Sec.~\ref{prot2}. For the shown computational example we have set system parameters with significant deviations from the ideal case, with  $J_2 T=\pi/2+0.18$, $J_1 T=\pi/2+0.14$, $j_1 T=0.06$, $j_2 T=2\pi+0.19$, $\Delta_2 T=\pi/2-0.24$, $\Delta_1 T=\pi/2+0.1$, $\delta_1 T=-0.04$, and $\delta_2 T=0.12$.  The lattice size is chosen to be $N=100$.}
	\label{result}
\end{figure*}

In the Majorana representation, the above six steps, as depicted in Fig.~\ref{braidp1}, result in the braiding transformation $\gamma_{0,1}^L\rightarrow -\gamma_{0,2}^L$ and $\gamma_{0,2}^L\rightarrow \gamma_{0,1}^L$, while leaving the other Majorana modes invariant. We have thus achieved the braiding unitary $U_L$ necessary to construct $P_L$ and $Z_L$ gates as claimed above. Figures~\ref{result}(a) and (b) depict computational examples via the evolution of Majorana correlation functions between the three involved Majorana modes in the protocol.
There, the initial state is chosen to be $|+\rangle=1/\sqrt{2}\left(|01\rangle +|10\rangle\right)$, so that $\langle\mathrm{i}\gamma_{0,1}^L\gamma_{0,1}^R\rangle=\langle\mathrm{i}\gamma_{0,2}^L\gamma_{0,2}^R\rangle=\langle\mathrm{i}\gamma_\pi^R\gamma_\pi^L\rangle=1$, with any other cross correlation functions being zero. The success of the protocol is signified by the change in the cross correlations $\langle \mathrm{i} \gamma_{0,1}^L \gamma_{0,2}^R\rangle $ and $\langle \mathrm{i} \gamma_{0,2}^L \gamma_{0,1}^R\rangle $, which become $1$ or $-1$ at the end of the protocol. The shown correlation functions in the computational example confirm the successful implementation of the braiding unitaries $U_L$ and $U_R$. It should be emphasized that the system parameters used in the computational example have been tuned away from the ideal case, so fine tuning of the system parameters is indeed unnecessary.

{  In  Fig.~\ref{result}(c), we plot the eigenphase spectrum of the two-period Floquet operator $\mathcal{U}^2$, where the eigenphase $\varepsilon_2$ satisfies $\mathcal{U}^2|\varepsilon_2\rangle = \exp\left(-\mathrm{i}\varepsilon_2 T\right)|\varepsilon_2\rangle$ for a given eigenstate $|\varepsilon_2\rangle$. In particular, it can be observed that a large quasienergy gap exists between the bulk and the zero edge states throughout the computation protocol. This spectral feature is necessary to ensure that adiabaticity condition may hold during the holonomic process. Indeed, we have checked that under the timescale used in our numerics, the diabatic error, which is obtained by projecting the final states onto the subspace spanned by the initial Majorana modes, is of order $10^{-4}$ or smaller.
	
	Apart from diabatic error, another source of error that may arise in the physical implementation of the aforementioned protocol is caused by the imperfection in tuning each adiabatic parameter $\phi$ perfectly from $0$ to $\pi/2$ at each step of the protocol. However, by realizing that the result of our protocol is determined by the solid angle formed by the holonomic path in the parameter space \cite{Tjun4,magic}, a sufficiently small error in the end points of the adiabatic parameter at each step of the protocol will only result in a small deformation of the holonomic path, which on average tends to preserve its resulting solid angle. As a result, our protocol at least enjoys the expected robustness characteristic of a holomonic computation protocol.} % adiabatic error < 10^{-4}}

%\begin{widetext}
\ \ \
%\end{widetext}

\subsection{Hadamard gate and Pauli $X$ gate}
\label{prot2}

\begin{figure*}
	\begin{center}
		\includegraphics[scale=0.8]{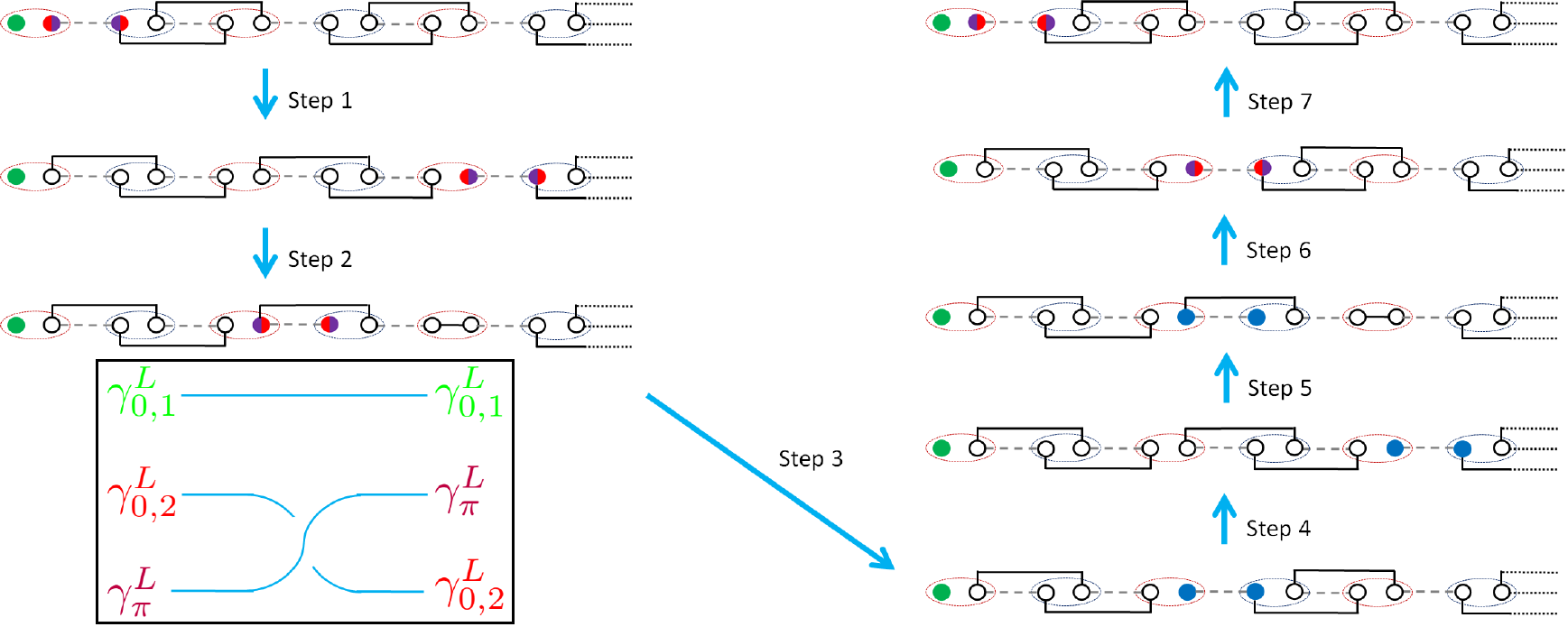}
	\end{center}
	\caption{(color online). Schematic of the holonomic protocol to braid $\gamma_\pi^L$ and $\gamma_{0,2}^L$. Only the first three lattice sites are shown. Blue coloured circles denote the two-period Majorana modes due to the superoosition of Majorana zero and $\pi$ modes. The meaning of the other symbols are the same as those in Fig.~\ref{braidp1}.}
	\label{braidp2}
\end{figure*}

Upon implementation of phase gate and Pauli $Z$ gate, {  we will now present the implementation of Hadamard gate ($\mathcal{H}$) and Pauli $X$ gate with another set of braiding operations, i.e., the braiding between $\gamma_\pi^L$ and $\gamma_{0,2}^L$ or between $\gamma_\pi^R$ and $\gamma_{0,1}^R$. It is again straightforward to verify, by using the encoding relations in Eq.~(\ref{qubits}), that $V_L =\exp\left[(\pi/4)\gamma_\pi^L \gamma_{0,2}^L\right]\equiv \mathcal{H}_L Z_L $, $X_L= V_L^2 Z_L=\exp\left[(\pi/2)\gamma_\pi^L \gamma_{0,1}^L\right]$, $V_R=\exp\left[(\pi/4)\gamma_\pi^R \gamma_{0,1}^R\right]\equiv \mathcal{H}_R Z_R$, and $X_R=V_R^2 Z_R =\exp\left[(\pi/2)\gamma_{0,2}^R \gamma_\pi^R\right]$. That is, the braiding unitary $V_s$ realizes the product of the Hadamard gate and the $Z$ gate, which can be further used to realize the $X$ gate by combining it with the Pauli $Z$ gate described in Sec.~\ref{prot1}.

In the following, we propose that braiding between $\gamma_\pi^L$ and $\gamma_{0,2}^L$ (and similarly between $\gamma_\pi^R$ and $\gamma_{0,1}^R$) can be accomplished in seven steps. Similar to the braiding procedure described earlier in Sec.~\ref{prot1}, each step amounts to adiabatically deforming the system Hamiltonian so as to move the Majorana modes around different lattice sites { (as depicted in Fig.~\ref{braidp2})}. Except for steps 3 and 6 below, such adiabatic deformation is characterized by the adiabatic parameter $\phi$ which is slowly varied at the beginning of each new period, such that it starts at $\phi=0$ and ends at $\phi=\pi/2$ at each step. In steps 3 and 6, we adopt a different adiabatic procedure, which follows a technique introduced earlier by us in Ref.~\cite{RG}. In these steps, we introduce a different adiabatic parameter $s$, which is tuned \emph{every other period}. This procedure amounts to creating a non-Abelian rotation in the subspace spanned by $\gamma_{0,2}^L$ and $\gamma_\pi^L$, which is possible due to the fact that these Majorana modes will now adiabatically follow the two-period Floquet operator $U^2$, which commutes with both $\gamma_{0,2}^L$ and $\gamma_\pi^L$. For simplicity, we will again present the steps of our protocol by focusing on the ideal case and leaving more technical details in Appendix~\ref{app3}.}

%In order to braid $\gamma_\pi^L$ and $\gamma_{0,2}^L$ with different quasienergy values, we adopt a technique introduced by us earlier in Ref.~\cite{RG}.  In short, both $\gamma_\pi^L$ and $\gamma_{0,2}^L$ modes can be regarded as topological zero edge modes of $\mathcal{U}^2$, namely, the Floquet operator associated with every two periods.  This insight suggests that if we introduce adiabatic deformation of the quantum wire Hamiltonian every other period, then it is possible to induce rotation in the ``degenerate" subspace formed by $\pi$ and zero edge modes. As before, here we only present our protocol to braid $\gamma_{0,2}^L$ and $\gamma_\pi^L$ in the ideal case. Braiding $\gamma_{0,1}^R$ and $\gamma_\pi^R$ can be likewise achieved by tuning instead the appropriate hopping and pairing strength near the right end of the quantum wire. At each step below, except for steps 3 and 6, the adiabatic parameter $\phi$ is slowly tuned at the beginning of each period from $0$ to $\pi/2$. We again leave more technical details in Appendix~\ref{app3}. %\textcolor{red}{why you can avoid the rotation between $\gamma_{0,1}^L$ and $\gamma_\pi^L$?}

\textit{Step 1.---} In this step, $\gamma_{0,2}^L$ and $\gamma_\pi^L$ are moved to the $n+1$-th lattice site. In order to reduce unwanted non-Abelian rotation between the two degenerate modes $\gamma_{0,1}^L$ and $\gamma_{0,2}^L$, it is better to take large $n>2$. Certainly the value of $n$ is also limited by the actual lattice size in order to avoid potential overlap with Majorana modes at the right edge. {  As detailed in Appendix~\ref{app3}}, we find that this step can be easily carried out by adiabatically tuning $\left(j_{\mathrm{inter},k}+\delta_{\mathrm{inter},k}\right) T=2\pi\cos\phi$ and $j_{\mathrm{intra},k}T=\delta_{\mathrm{intra},k} T=\pi\sin\phi$, with $\left(j_{\mathrm{inter},k}-\delta_{\mathrm{inter},k}\right) T=2\pi$, where $k=1,2,\cdots, n$. This results in $\gamma_\pi^L=\frac{1}{\sqrt{2}} \left(\gamma_{A,n+1}^\beta +\gamma_{B,n+1}^\alpha\right)$ and $\gamma_{0,2}^L=\frac{1}{\sqrt{2}} \left(\gamma_{A,n+1}^\beta -\gamma_{B,n+1}^\alpha\right)$ at the end of this step, both can be sufficiently away from the other zero mode $\gamma_{0,1}^L$ on the left edge.

\textit{Step 2.---} In this step, we move $\gamma_{0,2}^L$ and $\gamma_\pi^L$ to the $n$th lattice site, while at the same time exchanging their superposition structure, i.e., $\gamma_{0,2}^L$ and $\gamma_\pi^L$ respectively become symmetric and antisymmetric superpositions of two Majorana operators. This is accomplished by adding a potential bias at sublattice $A$ in the $(n+1)$th lattice site with strength $VT=2\pi \sin\phi$, such that $H_1=\cdots+Vc_{A,n+1}^\dagger c_{A,n+1}$, and further tuning $j_{\mathrm{inter},n}T =-\delta_{\mathrm{inter},n}T = \pi\cos\phi$, so that  $\gamma_\pi^L=\frac{1}{\sqrt{2}}  \left(\gamma_{B,n}^\alpha- \gamma_{A,n}^\beta \right)$ and $\gamma_{0,2}^L=  \frac{1}{\sqrt{2}}\left(\gamma_{A,n}^\beta +\gamma_{B,n}^\alpha \right)$ at the end of the step.

\textit{Step 3.---}{  As outlined before, this step amounts to introducing a non-Abelian rotation in the subspace spanned by zero and $\pi$ edge modes, both regarded as zero modes of $\mathcal{U}^2$. This is accomplished by varying $VT=\pi\left(1-f(s)\right)$, $j_{\mathrm{intra},n} T=\delta_{\mathrm{intra},n} T= \frac{\pi}{2}\left(1-f(s)\right)$, and $j_{\mathrm{inter},n} T=\pi \left(1+f(s)\right)$, where $VT$ is the potential bias introduced in step 2, $f(s)$ is a rather arbitrary function which increases from $-1$ to $1$ as the adiabatic parameter $s$ is adiabatically tuned every other period. While difficult to solve analytically, we have numerically verified that at the end of the step, $\gamma_\pi^L = \gamma_{A,n}^\beta $ and $\gamma_{0,2}^L = -\gamma_{B,n}^\alpha $. It should be noted that while we keep the same notations as before, $\gamma_\pi^L$ and $\gamma_{0,2}^L$ are no longer Majorana $\pi$ and zero modes with respect to $\mathcal{U}$, but they are still Majorana zero modes of $\mathcal{U}^2$ \cite{RG}.}

\textit{Step 4.---}We further tune the system according to $\left(j_{\mathrm{inter},n}+\delta_{\mathrm{inter},n}\right) T=2\pi\cos\phi$, $\left(j_{\mathrm{inter},n}-\delta_{\mathrm{inter},n}\right) T=2\pi$, and $j_{\mathrm{intra},n} T=\delta_{\mathrm{intra},n} T=\pi\sin\phi$. This results in moving $\gamma_\pi^L$ and $\gamma_{0,2}^L$ to $\gamma_{A,n+1}^\beta$ and $-\gamma_{B,n+1}^\alpha$ respectively.

\textit{Step 5.---}{ This step is identical to step 2  in terms of Hamiltonian manipulation, and it now moves $\gamma_\pi^L$ and $\gamma_{0,2}^L$ to $\gamma_{B,n}^\alpha$ and $\gamma_{A,n}^\beta$ respectively.}

\textit{Step 6.---}{  This step is identical to step 3. Namely, the system parameters are parameterized by the adiabatic parameter $s$ as described in step 3, which is only tuned every other period. Because $\gamma_\pi^L$ and $\gamma_{0,2}^L$ are already superposition Majorana zero and $\pi$ modes, our numeric shows that they transform as $\gamma_\pi^L = \frac{1}{\sqrt{2}}\left(\gamma_{A,n}^\beta -\gamma_{B,n}^\alpha \right)$ and $\gamma_{0,2}^L=  -\frac{1}{\sqrt{2}}\left(\gamma_{B,n}^\alpha +\gamma_{A,n}^\beta \right)$ at the end of the step. That is, $\gamma_\pi^L$ and $\gamma_{0,2}^L$ are now respectively Majorana zero and $\pi$ modes of $\mathcal{U}$.}

\textit{Step 7.---}As the final step, we need to return the Hamiltonian to its original form. This is done by tuning $\left(j_{\mathrm{inter},k}+\delta_{\mathrm{inter},k}\right) T=2\pi\sin\phi$, $\left(j_{\mathrm{inter},k}-\delta_{\mathrm{inter},k}\right) T=2\pi$, and $j_{\mathrm{intra},k}T=\delta_{\mathrm{intra},k} T=\pi\cos\phi$, where $k=1,2,\cdots, n$. This step also moves $\gamma_\pi^L$ and $\gamma_{0,2}^L$ back to the first site. {  At the end of the step, we find that $\gamma_\pi^L\rightarrow \frac{1}{\sqrt{2}}\left(\gamma_{A,1}^\beta -\gamma_{B,1}^\alpha \right)$ (which is the initial $\gamma_{0,2}^L$) and $\gamma_{0,2}^L\rightarrow   -\frac{1}{\sqrt{2}}\left(\gamma_{A,1}^\beta +\gamma_{B,1}^\alpha \right)$ (which is the initial $\gamma_\pi^L$ multiplied by $-1$), which completes the braiding operation}.

%\begin{widetext}
\ \ \
%\end{widetext}

The seven steps above are schematically depicted in Fig.~\ref{braidp2}, with the net outcome $\gamma_{0,2}^L\rightarrow-\gamma_\pi^L$ and $\gamma_\pi^L\rightarrow \gamma_{0,2}^L$. {  Even with system parameters slightly deviating from the ideal values, our numerical results show that the aforementioned protocol still yields the desired braiding operation with a very good fidelity, as summarized in Figs.~\ref{result}(d) and (e). There, we take the same initial state and parameter values as those in Sec.~\ref{prot1}, $n=4$ in step 1, and $f(s)=\cos\left(s\pi\right)$ in step 3 and step 6, where $s$ decreases slowly every other period from $1$ to $0$.} The success of the protocol is signified by the change in cross correlations $\langle \mathrm{i} \gamma_\pi^R \gamma_{0,2}^L\rangle$ and $\langle \mathrm{i} \gamma_\pi^L \gamma_{0,2}^R\rangle$ ($\langle \mathrm{i} \gamma_{0,1}^R \gamma_\pi^L\rangle$ and $\langle \mathrm{i} \gamma_{0,1}^L \gamma_\pi^R\rangle$) to $1$ or $-1$ for braiding between $\gamma_{0,2}^L$ and $\gamma_\pi^L$ ($\gamma_{0,1}^R$ and $\gamma_\pi^R$).

 {  In Fig.~\ref{result}(f), we have plotted the eigenphase spectrum of $\mathcal{U}^2$ throughout the whole process. In particular, it confirms that zero and $\pi$ edge modes maintain a large quasienergy gap from the instantaneous bulk states throughout the seven steps of adiabatic manipulation, which is necessary to ensure that adiabaticity condition remains valid in our protocol. Indeed, under the timescale used in our numerics, the diabatic error is found to be very small, i.e., $<10^{-4}$. Apart from the large bulk gap, it is also necessary for the eigenphase spectrum to maintain very small quasienergy splitting between the Majorana modes, so as to ensure that all Majorana modes remain degenerate with one another and there is no accidental qubit readout throughout the protocol. The former is also especially important in steps 3 and 6 of our protocol to ensure that the non-Abelian rotation between Majorana zero and $\pi$ modes arises solely due to geometrical and not dynamical effect. For these reasons, we have also checked numerically that the quasienergy splitting of the Majorana modes throughout all the steps in the whole protocol is of order $10^{-7}$ or smaller.
 	
 Finally, since Majorana zero and $\pi$ modes become effectively degenerate during steps 3 and 6 of our protocol due to the nature of our adiabatic manipulation, one may wonder if our system becomes more susceptible to errors due to steps 3 and 6. Put another way,
 will the two Majorana modes hybridize easily during our adiabatic protocol?  To address this important question, we first note that perturbations capable of hybridizing zero and $\pi$ modes must have $2T$ periodicity. This requirement is incompatible with the periodicity of the Hamiltonian in the absence of the adiabatic manipulation. In steps 3 and 6 of our protocol, our adiabatic manipulation amounts to only tuning the system parameters that are always modulated at a period of $T$. As a consequence, our manipulation itself is not a dangerous $2T$ periodic perturbation to hybridize the two Majorana modes. Thus, the main source of errors still comes from the imperfection of Hamiltonian manipulation. } % Splitting in edge modes during step 3 and 6 is <10^{-7}

\subsection{Qubit readout}
\label{prot3}

The last step in a typical quantum computation task is to readout qubits, which allows one to confirm that a sequence of gate operations applied on an input qubit indeed gives the intended outcome. Our system uses three physical qubits to encode two logical qubits. As elucidated in Sec.~\ref{encode}, two of these three physical qubits originate from the chiral symmetry protected edge states at both ends of the lattice. By systematically introducing a chiral symmetry breaking term in the Hamiltonian, the degeneracy of these two edge states can then be lifted, which thus allows one to distinguish between the four logical-qubit states according to their quasienergy values.

To be more explicit, we may add the following symmetry-breaking terms to the Hamiltonian in Eq.~(\ref{model}),

\begin{equation}
H_{\rm break}= \sum_i \left[\left(\mu_1 + \mu_2\right)c_{A,i}^\dagger c_{A,i}+\left(\mu_1 - \mu_2\right)c_{B,i}^\dagger c_{B,i}\right] \;.
\end{equation}

\begin{figure}
	\begin{center}
		\includegraphics[scale=0.3]{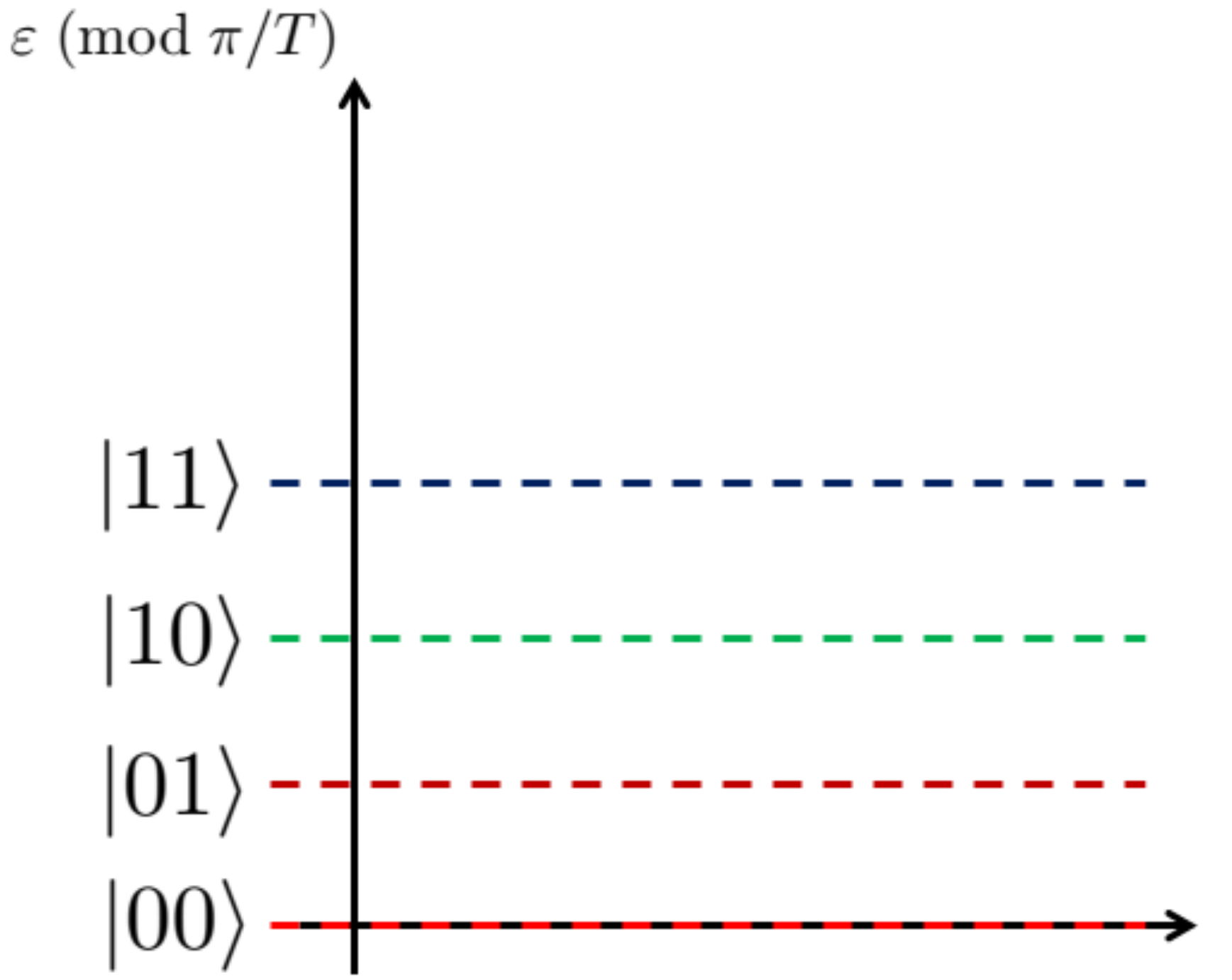}
	\end{center}
	\caption{(color online). The four qubit states can be uniquely distinguished by introducing a chiral symmetry breaking term in the Hamiltonian, which lifts all degeneracy in their quasienergy values.}
	\label{read}
\end{figure}

\noindent It can be easily verified that $H_{\rm break}$ violates the chiral symmetry defined in Sec.~\ref{sym}. In particular, $\mu_1$ shifts the quasienergy of both edge states by an equal amount. As a result, qubit states associated with occupied edge states, such as $|01\rangle$, $|10\rangle$, and $|11\rangle$, will have different quasienergies (modulo $\pi/T$) as compared with $|00\rangle$, which has neither fermionic nor Majorana excitations. Moreover, $|11\rangle$ will have different quasienergies (modulo $\pi/T$) as compared with $|01\rangle$ and $|10\rangle$ since the former has both edge states occupied. Finally, $\mu_2$ introduces a quasienergy difference between the two edge states, which results in $|01\rangle$ and $|10\rangle$ having different quasienergy values. Thus, in the presence of $H_{\rm break}$, all four qubit states now have different quasienergy values (modulo $\pi/T$), as illustrated in Fig.~\ref{read}. In practice, the difference in quasienergy can be indirectly probed by, for example,  irradiating the system with electromagnetic waves, which results in qubit-state dependent resonant frequency \cite{probe1,probe2,probe3}.

\subsection{Implementation of simple quantum algorithms}
\label{algorithm}

\begin{figure*}
	\begin{center}
		\includegraphics[scale=0.9]{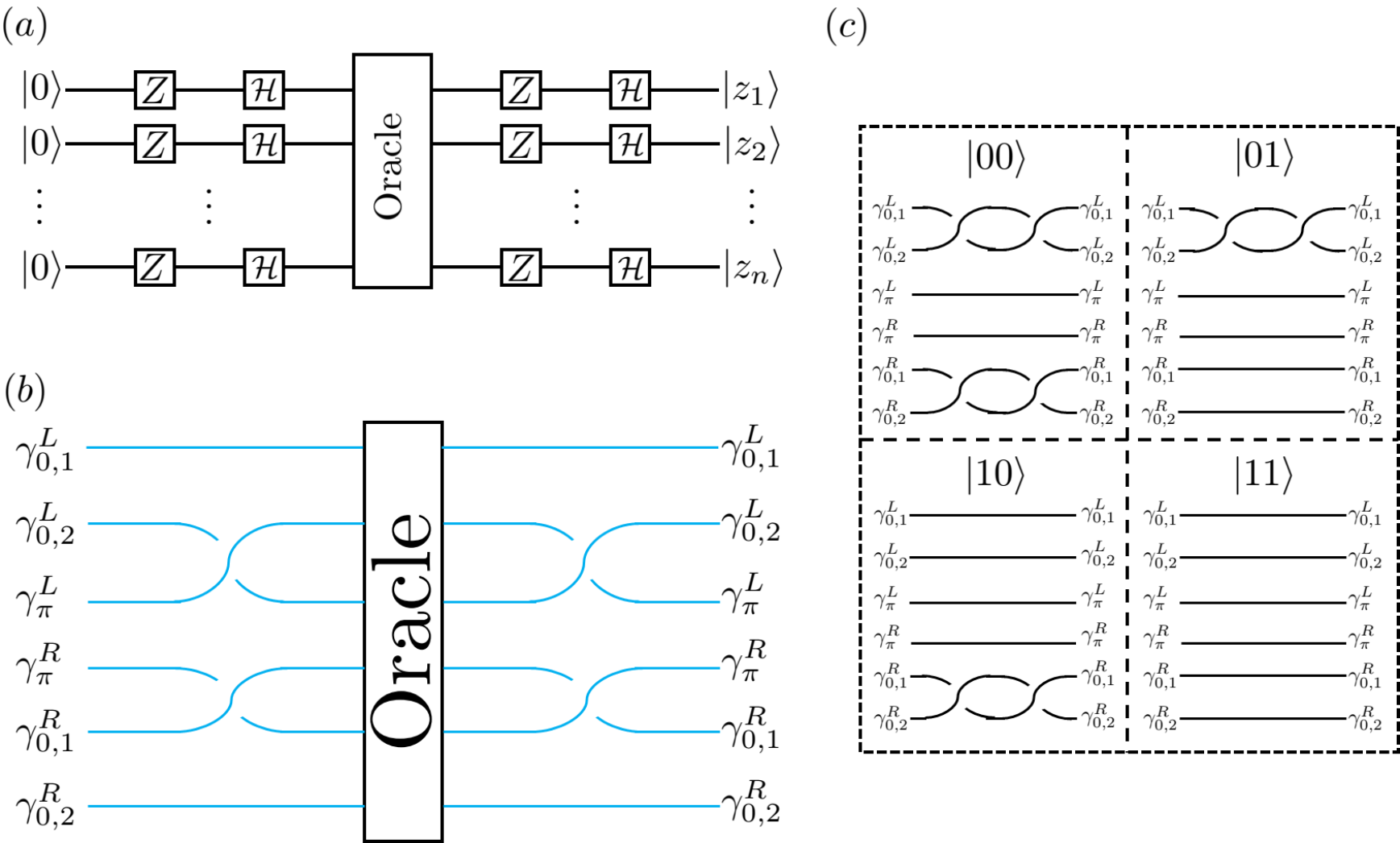}
	\end{center}
	\caption{(a) Description of our quantum search algorithm with $n$ qubits. (b) Implementation of the algorithm in our system with two logical qubits. (c)   {The four different choices for the associated oracle operators in panel (b).}}
	\label{oracle}
\end{figure*}

To demonstrate the application of our results presented in Sec.~\ref{prot1} to \ref{prot3}, we now illustrate two simple quantum algorithms realized by the gate operations developed in Sec.~\ref{prot1} and \ref{prot2}. { The first one is a simple inversion algorithm, which can be viewed as a simplified version of the Grover's search algorithm \cite{grover}}. As compared with the latter, our algorithm assumes a special structure of a database which maps a number $z\in \left\lbrace 1,2\cdots, 2^n\right\rbrace$ to $\bar{z}=2^n-z$. In other words, one needs to obtain $z$, given $\bar{z}$, {quantum mechanically}. By employing the quantum circuit in Fig.~\ref{oracle}(a), where the oracle operator is to be defined below, this can be accomplished in just a single step, {similar to its classical counterpart. While it does not demonstrate the advantage of quantum over classical computation, this simple example illustrates how quantum computation works.}

To be more explicit, let $\vec{z}=\left(z_1,\cdots , z_n\right)$ be a column vector representing the binary expansion of $z$, i.e., $z=z_1\times 2^0+\cdots +z_n \times  2^{n-1}$, and define $|\vec{z}\rangle =|z_1\cdots z_n\rangle$. Next, define the oracle operator as $\mathcal{O}=\prod_{i=1}^n Z_i^{\bar{z}_i}$, where $Z_i$ is the Pauli $Z$ gate acting on qubit $i$, $\bar{z}_i=z_i\oplus 1$, and $\oplus$ is the addition operation modulo two. It is now straightforward to show that Fig.~\ref{oracle}(a) indeed maps an input $|\vec{0}\rangle$ to the desired output $|\vec{z}\rangle$,

%\begin{widetext}
\ \ \
%\end{widetext}

\begin{eqnarray}
|\vec{0}\rangle &\xrightarrow{\left(\mathcal{H}Z\right)^{\bigotimes n}}& \sum_{\vec{x}} |\vec{x}\rangle \nonumber \\
&\xrightarrow{\;\;\;\;\;\mathcal{O}\;\;\;\;\;\;}& \sum_{\vec{x}} (-1)^{\vec{x}\cdot \vec{\bar{z}}} |\vec{x}\rangle \nonumber \\
 &\xrightarrow{\left(\mathcal{H}Z\right)^{\bigotimes n}}& \sum_{\vec{x},\vec{y}} (-1)^{\vec{x}\cdot \left(\vec{z}-\vec{y}\right)} |\vec{y}\rangle \nonumber \\
 &=& \sum_{\vec{y}} \delta_{\vec{z},\vec{y}} |\vec{y} \rangle \nonumber \\
 &=& |\vec{z} \rangle \;,
 \label{or}
\end{eqnarray}

\n where we have suppressed any normalization constant for brevity, $\vec{x}\cdot \vec{y}=x_1y_1\oplus\cdots \oplus x_n y_n$, and we have used the fact that $\sum_{\vec{x}} (-1)^{\vec{x}\cdot\vec{y}}=\delta_{\vec{0},\vec{y}}$.

To implement the above algorithm in our system, we first note that a single superlattice is already capable of hosting two logical qubits, and the two gate operations above, i.e., the $Z$ and $\mathcal{H}Z$ gates, can be implemented by braiding Majorana modes according to the protocols outlined in Sec.~\ref{prot1} and \ref{prot2} respectively. In the two-qubit case, our algorithm is capable of finding an object from a database of size $2^2=4$. In terms of braiding operations, our circuit and its associated oracle operator are depicted in Figs.~\ref{oracle}(b) and (c). Assuming that all Majorana modes are initialized in $|00\rangle$ state, protocol described in Sec.~\ref{prot2} is first carried out to implement $\mathcal{H}_LZ_L$ and $\mathcal{H}_R Z_R$ gate operations, which brings our qubit state to an equal-weight superposition of all qubit basis states. Next, depending on the input we supply to the black box, the oracle operator will execute one of the four sets of Pauli $Z_L$ and $Z_R$ gates as illustrated in Fig.~\ref{oracle}, all of which are achievable through the protocol developed in Sec.~\ref{prot1}. This flips the sign of the weight of some qubit basis states. Lastly, another $\mathcal{H}_L Z_L$ and $\mathcal{H}_R Z_R$ gates are applied to bring our qubit state to the desired output. This output is then measured by implementing the readout process described in Sec.~\ref{read}.

 It can be seen that the same oracle can be used to implement the Deutsch-Jozsa algorithm \cite{Deutsch}, capable of identifying whether a particular function is constant, i.e., $g(x)=0$ (or $g(x)=1$) for any input $x\in \left\lbrace 1,\cdots , 2^n\right\rbrace$, or balanced, i.e., $g(x)=0$ for half the inputs and $g(x)=0$  for the other half.  To proceed, note that any balanced or constant function can be expressed as $g(x)=\vec{x}\cdot \vec{z}\oplus k$ for a fixed but unknown $z\in \left\lbrace 1,\cdots , 2^n\right\rbrace $ and $k=0,1$. Indeed, it can be checked that $g(x)$ is constant if and only if $\vec{z}=\vec{0}$, otherwise it is balanced. Therefore, Deutsch-Jozsa algorithm proceeds in the same way as above, i.e., as depicted in Fig.~\ref{oracle}(a)-(c), with $\vec{x}\cdot\vec{\bar{z}}$ being now identified as the function $g(x)$. The latter being constant is thus identified when $|1\cdots 1\rangle$ appears as output; any other output implies $g(x)$ being balanced.  In fact, similar braiding-based oracle has also been used in Ref.~\cite{wire} for exactly this purpose, although a minimum of three wires is required to construct an oracle of size $N=2^2=4$ in the setup of Ref.~\cite{wire}.  By contrast, here we only require a single wire after exploiting the coexistence of two pairs of MZMs and one pair of Majorana $\pi$ modes.

\subsection{Scalability and implementation of entangling gates}
\label{scalability}

\begin{figure*}
	\begin{center}
		\includegraphics[scale=0.9]{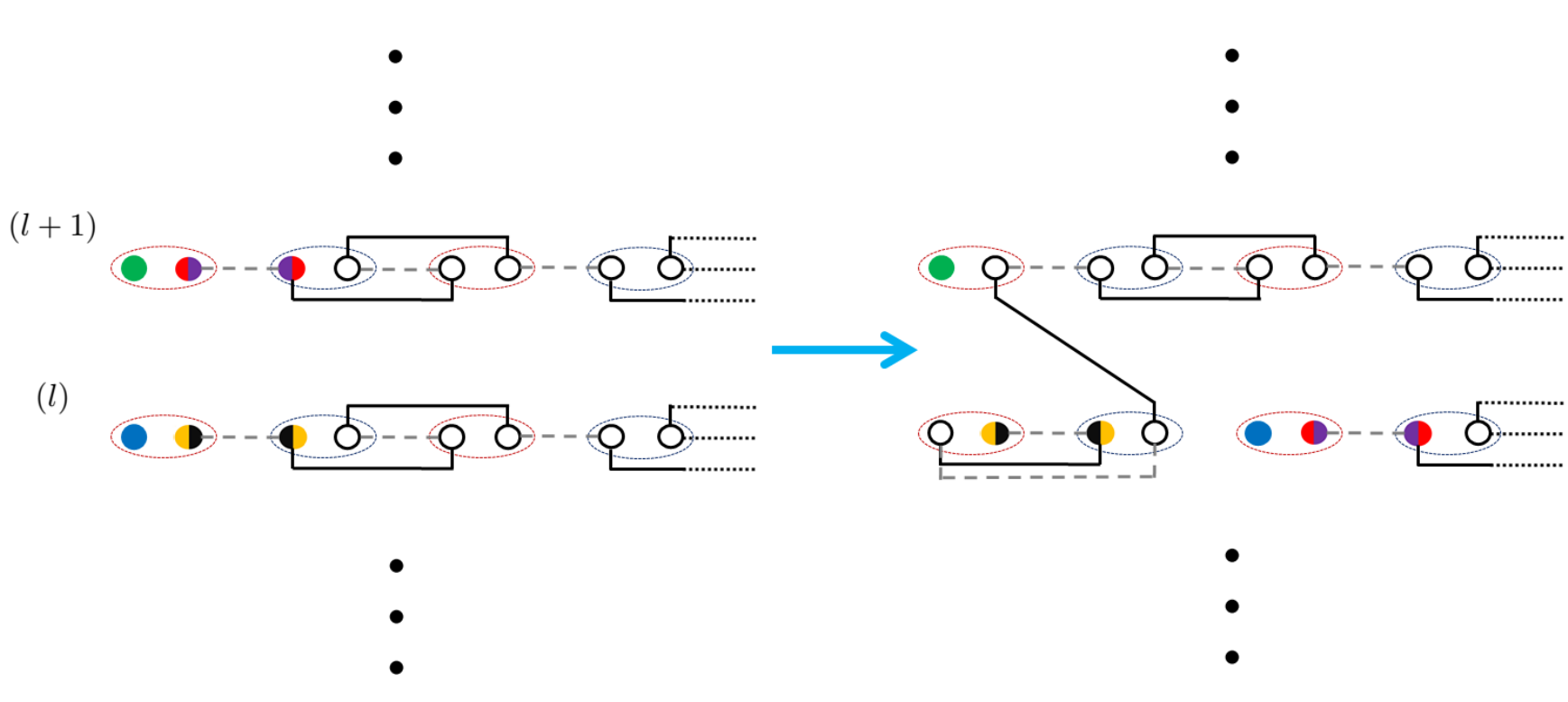}
	\end{center}
	\caption{Generalization of our single-wire braiding scheme to an array of wires. Majorana modes can be moved to another site belonging to the same (blue circle) or different wires (red-magenta circle) by appropriately tuning intra- and inter-wire hopping and pairing strengths.} %Ideal parameter values are assumed for simplicity.}
	\label{scale}
\end{figure*}

Given that two logical qubits are encoded and manipulated in a 1D setup, it is important to examine the possibility of scaling up our proposal.  There are two routes to scale up. The first route is to consider many zero modes and $\pi$ modes in one single quantum wire. In principle, their coexistence can be used to encode multi-qubit quantum information and it is not hard to imagine that certain quantum information processing becomes possible. This is an exciting target but we yet need to investigate how to braid two particular edge modes out of many without affecting the rest.  The other route for scaling up is to add more wires arranged in parallel with each other, as shown in Fig.~\ref{scale}.  Edge modes belonging to different wires can also be braided by turning on hopping and/or pairing between the wires.  The actual braiding protocols between two such Majorana modes from different wires can be designed by slightly modifying the protocols introduced in Sec.~\ref{prot1} and Sec.~\ref{prot2}. For example, braiding Majorana modes marked by blue and red circles in Fig.~\ref{scale} can be obtained by directly applying the protocol of Sec.~\ref{prot1} on wire labelled $(l)$, with step 2 and step 6 being slightly modified by introducing interwire hopping and pairing in order to move two Majorana modes from wire $(l+1)$ to wire $(l)$, as shown in Fig.~\ref{scale}.

%\begin{widetext}
\ \ \
%\end{widetext}

As a promising side finding, in the following we show that by considering only the two wires $(l)$ and $(l+1)$ illustrated in Fig.~\ref{scale}, entangling gates such as CNOT and other controlled-Pauli gates can be implemented through a series of braiding and measurement operations only. For brevity, we will only present the construction of a CNOT gate with the first and second qubits being the target and control qubits respectively, encoded in wire $(l)$, with its Majorana modes denoted as $\gamma_{0,1}^{(l),s}$, $\gamma_{0,2}^{(l),s}$, and $\gamma_\pi^{(l),s}$, where $s=L,R$. The additional six Majorana modes in wire $(l+1)$ give rise to additional three logical qubits, but for the purpose of implementing controlled-Pauli gates, only a single qubit encoded by $\gamma_{0,1}^{(l+1),L}$ and $\gamma_{0,2}^{(l+1),L}$ will be used as ancilla, whereas the other two qubits can be used as additional stabilizer operators. It is further assumed that the ancilla is prepared in $|1\rangle_a$, which can be done by following the protocol of Sec.~\ref{prot3}.

We start by writing the CNOT unitary as $U(X_L)=\exp\left[\mathrm{i}\pi/4\left(1-Z_R\right)\left(1-X_L\right)\right]$, which can be written in terms of Majorana modes as

\begin{eqnarray}
U(X_L)&=& \exp\left[\mathrm{i}\pi/4\right]\times \exp\left[\mathrm{i}\pi/4\left(\gamma_{0,1}^{(l),R}\gamma_{0,2}^{(l),R}\gamma_{0,1}^{(l),L}\gamma_\pi^{(l),L}\right)\right] \nonumber \\
&& \times \exp\left[\pi/4 \gamma_{0,1}^{(l),R}\gamma_{0,2}^{(l),R} \right]\times \exp\left[\pi/4\gamma_\pi^{(l),L}\gamma_{0,1}^{(l),L} \right]\;. \nonumber \\
\end{eqnarray}

\n The third and fourth exponentials of $U(X_L)$ are simply the braiding unitaries discussed in Sec.~\ref{prot1} and Sec.~\ref{prot2}. On the other hand, the second exponential can be implemented by performing projective measurements on $\Pi_1=\gamma_{0,1}^{(l),R}\gamma_{0,2}^{(l),R}\gamma_\pi^{(l),L}\gamma_{0,1}^{(l+1),L}$ and $\Pi_2=\mathrm{i}\gamma_{0,1}^{(l+1),L} \gamma_{0,1}^{(l),L}$, followed by measurement dependent corrections, which are realizable through braiding \cite{flux, magic3}.

{To be more explicit, we can write $\Pi_1=\frac{1}{2}(1+p_1\gamma_{0,1}^{(l),R}\gamma_{0,2}^{(l),R}\gamma_\pi^{(l),L}\gamma_{0,1}^{(l+1),L})$ and $\Pi_2=\frac{1}{2}(1+p_2 \mathrm{i}\gamma_{0,1}^{(l+1),L} \gamma_{0,1}^{(l),L})$, where $p_1,p_2=\pm 1$ are the measurement results of $\Pi_1$ and $\Pi_2$ respectively. The effect of the two measurements can then be written as
	
\begin{eqnarray}
\Pi_2 \Pi_1 &=& \frac{1}{4}\left(1-\mathrm{i} p_1 \gamma_{0,1}^{(l),R} \gamma_{0,2}^{(l),R} \gamma_\pi^{(l),L} \gamma_{0,2}^{(l+1),L}+p_2 \gamma_{0,2}^{(l+1),L}\gamma_{0,1}^{(l),L}\right. \nonumber \\
&& +\left.p_1p_2 \mathrm{i} \gamma_{0,1}^{(l),R}\gamma_{0,2}^{(l),R}\gamma_{0,1}^{(l),L} \gamma_\pi^{(l),L}\right) \;, \nonumber \\
&=& \frac{1}{4}\left(1-\mathrm{i} p_1 \gamma_{0,1}^{(l),R}\gamma_{0,2}^{(l),R} \gamma_\pi^{(l),L}\gamma_{0,2}^{(l+1),L}\right) \nonumber \\
&& \times \left(1+p_2 \gamma_{0,2}^{(l+1),L}\gamma_{0,1}^{(l),L}\right) \;, \label{intermed}
\end{eqnarray}	

\n where we have used $\mathrm{i}\gamma_{0,1}^{(l+1),L}\gamma_{0,2}^{(l+1),L}|1\rangle_a=-|1\rangle_a$. By further applying $U_1(p_2)=\exp\left[-\frac{\pi}{4}p_2 \gamma_{0,2}^{(l+1),L}\gamma_{0,1}^{(l),L}\right]$, Eq.~(\ref{intermed}) becomes

\begin{equation}
U_1(p_2) \Pi_2 \Pi_1= \frac{1}{2\sqrt{2} } \left(1+\mathrm{i} p_1 p_2 \gamma_{0,1}^{(l),R}\gamma_{0,2}^{(l),R} \gamma_{0,1}^{(l),L} \gamma_\pi^{(l),L}\right) \;. \label{intermed2}
\end{equation}

\n Note that Eq.~(\ref{intermed2}) is equal to the second exponential of $U(X_L)$, up to a constant, provided $p_1p_2=1$. If $p_1p_2= -1$, further unitary $U_2=\exp\left[\frac{\pi}{2} \gamma_{0,1}^{(l),R}\gamma_{0,2}^{(l),R}\right]\times \exp\left[\frac{\pi}{2} \gamma_{0,1}^{(l),L} \gamma_\pi^{(l),L}\right]$ is applied to Eq.~(\ref{intermed2}), which leads also to the desired result.}

In our system, $\Pi_1$ can be carried out by first braiding $\gamma_\pi^{(l),L}$ and $\gamma_{0,2}^{(l+1),L}$, measuring $\Pi_1'=\gamma_{0,1}^{(l),R}\gamma_{0,2}^{(l),R}\gamma_{0,2}^{(l+1),L}\gamma_{0,1}^{(l+1),L}$ via the introduction of chiral symmetry breaking terms on the left half of wires $(l)$ and $(l+1)$, then finally undoing the braiding between $\gamma_\pi^{(l),L}$ and $\gamma_{0,2}^{(l+1),L}$. Likewise, $\Pi_2$ is carried out by first braiding $\gamma_{0,1}^{(l),L}$ and $\gamma_{0,2}^{(l+1),L}$, measuring $\Pi_2'=\mathrm{i} \gamma_{0,1}^{(l+1),L}\gamma_{0,2}^{(l+1),L}$ by introducing chiral symmetry breaking terms on wire $(l+1)$, then undoing the braiding between $\gamma_{0,1}^{(l),L}$ and $\gamma_{0,2}^{(l+1),L}$. After some algebra, $U(X_L)$ can finally be expressed as

\begin{widetext}

\begin{eqnarray}
U(X_L) &=& 2\exp\left[\mathrm{i}\pi/4\left(2-p_1p_2\right)\right]\times \exp\left[\pi/4\left(2-p_1p_2\right)\gamma_{0,1}^{(l),R} \gamma_{0,2}^{(l),R}\right] \times  \exp\left[\pi/4 p_1p_2 \gamma_\pi^{(l),L} \gamma_{0,1}^{(l),L}\right] \nonumber \\
&& \times  \exp\left[\pi/4 (p_2-1) \gamma_{0,1}^{(l),L}\gamma_{0,2}^{(l+1),L}\right] \times \Pi_2'  \times  \exp\left[\pi/4 \gamma_{0,1}^{(l),L}\gamma_{0,2}^{(l+1),L}\right] \times \exp\left[\pi/4 \gamma_{0,2}^{(l+1),L} \gamma_\pi^{(l),L}\right] \nonumber \\
&& \times \Pi_1' \times \exp\left[\pi/4 \gamma_\pi^{(l),L} \gamma_{0,2}^{(l+1),L} \right] \;, \label{cnot}
\end{eqnarray}

\end{widetext}

\n where $p_1,p_2=\pm 1$ are now the measurement results of $\Pi_1'$ and $\Pi_2'$ respectively. Other controlled-Pauli gates $U(P_L)=\exp\left[\mathrm{i}\pi/4\left(1-Z_R\right)\left(1-P_L\right)\right]$ can be implemented similarly, as $P_L$ can be expressed as a product of two Majorana modes.

\section{Discussion}
\label{discussion}

\subsection{Experimental consideration}
\label{exper}

Similar to other topological superconducting wires, {  it is expected that our model Eq.~(\ref{model}) can be potentially engineered in either cold-atom \cite{opt} or proximitized semi-conductor \cite{semi1,semi2} platforms, although such implementations may not be straightforward.} In a cold-atom setup, such a 1D model is formed by embedding optically trapped fermions inside a three dimensional Bose-Einstein condensate (BEC). The hopping and pairing terms are provided respectively by the two Raman lasers forming the optical lattice and the radio frequency (rf) field coupling the fermions with the surrounding BEC reservoir \cite{opt}. In this context the pairing and the hopping are in principle highly controllable. Sublattice degree of freedom can then be realized by using spatially periodic Raman lasers and rf field, which then allow two adjacent fermions to experience different hopping and pairing strength. Manipulation of the hopping and pairing strength to carry out the protocols described in Sec.~\ref{prot1} and Sec.~\ref{prot2} should be feasible by tuning the Rabi frequencies of the Raman lasers and rf field respectively. {In particular, switching between real and imaginary hopping and pairing parameters, i.e., between $H_1$ and $H_2$, can be done through switching between real and imaginary Rabi frequencies, which can be realized by appropriately setting the electric field profiles of the Raman lasers and rf field. Alternatively, by fixing the electric field profiles of the Raman lasers and rf field, one could also switch the phase of the hopping and pairing parameters by rapidly shaking the optical lattice at every integer multiple of $T/2$ \cite{comphop}.}

Following the discussion of Ref.~\cite{opt}, the coherence time-scale of Majorana modes in such cold atom setup can be extendable to the order of seconds. Meanwhile, given that the system parameters can be of the order of tens of kHz \cite{opt}, a single period of the system is typically of the order of $0.1$ ms so as to achieve the parameter regime in which SSH- and Kitaev-like edge states coexist. As shown in Fig.~\ref{result}, our braiding protocols are typically completed within $1000-2000$ periods to ensure adiabaticity. As a result, the quantum algorithms described in Sec.~\ref{algorithm} may take up to $0.8$ s to complete, provided that each gate operation on the first and second qubits are applied simultaneously, which is possible since two such gate operations require braiding between two left Majorana modes or two right Majorana modes only.

In proximitized semiconductor setup, topological superconductors are constructed by proximitizing 1D semiconducting wires with conventional $s$-wave superconductors \cite{semi1,semi2}. In addition, the wire is assumed to have a sufficiently large spin-orbit coupling and external magnetic field so as to open a gap in the vicinity of the crossing between the two spin-orbit bands. The proximitized $s$-wave superconductivity will then induce an effective $p$-wave pairing necessary for the creation of topological superconductors. {  In such a setup, however, our model might be more difficult to realize due to the necessity to switch between real and imaginary couplings. Indeed, even realizing imaginary hopping alone is already challenging in this setup. A plausible way to simulate our model in this setup might be to follow the proposal of Ref.~\cite{Tjun2,flux} through the use of Cooper pair box. In particular, the latter enables coupling between a pair of Majorana operators to be addressed directly, thus circumventing the need to realize imaginary hopping. In Appendix~\ref{app4}, we elucidate in detail the possibility of such Cooper pair boxes to realize Eq.~(\ref{model}). However, since designing an array of Cooper pair boxes to realize our model may take up some space and a number of wires to simulate all the Majorana operators in Eq.~(\ref{H}), it hinders the main purpose of our proposal to realize qubits in a minimal one dimensional setup. Therefore, while the use of Cooper pair boxes might be a good way to verify how our proposal works in experiment, it might not be a good platform to scale up our model for possible real life quantum computation applications. An alternative realization of our model, or at least a similar model which captures the main features of our model (coexistence of zero and $\pi$ edge modes belonging to different SPT phases) in semiconductor setup thus remains an interesting open question and is left for future studies.

Assuming that our model can eventually be implemented in such a semiconductor-superconductor setup, we will now compare the time-scale required to complete our braiding operations with the typical coherence lifetime of the system.} In particular, the coherence lifetime of Majorana modes in such a setup has been extensively studied \cite{lt1,lt2,lt3,lt4}, with estimates ranging between the order of tens of nanoseconds \cite{lt1} at worst to $>1$ min at best \cite{lt2}. On the other hand, typical energy scale in such a setup is of the order of $0.1$ meV (tens of GHz). As such, a single period of our system should be of the order of $0.1$ ns, and the time needed to complete the above quantum algorithms via our braiding protocols may be of the order of hundreds of nanoseconds, which in some cases may not exceed the coherence lifetime.

\subsection{Comparison with TQC}
\label{compare}

At first sight, our holonomic braiding-based protocols to realize quantum gate operations are very similar to typical approach in TQC.  Though TQC is also usually implemented through adiabatic holonomy, {  there are two main differences between TQC and our HQC, which are elucidated in detail below.
	
	In TQC, the qubits are encoded nonlocally, such as by using a pair of Majorana modes that are spatially separated, and are thus protected by any local perturbations. In our approach, the qubits are encoded both locally (through the occupation of the SSH edge states) and nonlocally (through the occupation of the nonlocal Kitaev fermion edge states). On the one hand, due to the local encoding of our qubits, our system loses the full topological protection typically offered in TQC due to the existence of certain local perturbations that may induce logical errors. On the other hand, since our qubits also require nonlocal encoding formed by the Majorana $\pi$ modes, together with the fact that Majorana $\pi$ modes and the SSH zero modes share some space together in the lattice, most dangerous local perturbations are forbidden by the total fermion parity symmetry of the system.
	
	An example of such a perturbation would be an onsite noise acting on one end of the system, which is capable of hybridizing two local Majorana zero modes and thus causing a logical $Z$ gate error. However, due to the existence of Majorana $\pi$ mode at each end of the lattice, the presence of such a perturbation would then also cause a parity flip of the associated nonlocal Kitaev fermion, which is thus incompatible with the conservation of total fermion parity. Hybridizing local zero modes in our system without flipping the parity of the nonlocal fermion thus requires either a very special local perturbation that will be very unlikely to take place or a nonlocal perturbation which involves adding onsite potential at both ends of the lattice simultaneously in the same spirit as the readout procedure described in Sec.~\ref{prot3}.

In terms of how gate operations are carried out, TQC usually requires that the non-Abelian Berry phase contribution in Eq.~(\ref{hol}) is zero during the holonomic cycle, so that the total geometric phase arises solely from the explicit monodromy \cite{geo1,geo2}. By writing
	
	\begin{eqnarray}
	\gamma_{0,a}^s&=&\sum_{D\in \left\lbrace A,B\right\rbrace,j\in\left\lbrace 1,2,\cdots N\right\rbrace,\nu\in\left\lbrace\alpha,\beta\right\rbrace} C_{a,\gamma_{D,j}^\nu}^s\gamma_{D,j}^\nu \;, \nonumber \\
	\gamma_\pi^s&=& \sum_{D\in \left\lbrace A,B\right\rbrace,j\in\left\lbrace 1,2,\cdots N\right\rbrace,\nu\in\left\lbrace\alpha,\beta\right\rbrace} C_{\pi,\gamma_{D,j}^\nu}^s\gamma_{D,j}^\nu \;,
	\end{eqnarray}

	 \n where $a\in\left\lbrace1,2\right\rbrace$ and $s\in\left\lbrace L,R\right\rbrace$, it can be verified that for all steps involved in Sec.~\ref{prot1} (at least in the ideal case), $\sum_{D,j,\nu} C_{a,\gamma_{D,j+1}^\nu}^s\frac{d}{d\phi} C_{b,\gamma_{D,j+1}^\nu}^s=0$, where $a,b\in\left\lbrace\pi,1,2\right\rbrace$. This implies that the protocol presented in Sec.~\ref{prot1} indeed contains no Berry phase contribution, and thus shares the same topological robustness as TQC in this aspect. Indeed, it can also be verified that replacing $\cos\phi$ ($\sin\phi$) with any function decreasing from $1$ to $0$ (increasing from $0$ to $1$) at each step in the protocol outlined in Sec.~\ref{prot1} does not change the net result.

On the other hand, the protocol elucidated in Sec.~\ref{prot2} would have also shared this topological robustness if not for its step 3 and step 6 processes.} In these two processes, non-Abelian Berry phase is necessarily introduced between $\gamma_\pi^L$ and $\gamma_{0,2}^L$ or between $\gamma_\pi^R$ and $\gamma_{0,1}^R$ to induce rotation between Majorana zero and $\pi$ modes. {However, we do not view this feature as a genuine weakness of our quantum computation protocols, because in actual physical implementation of the braiding of Majorana modes in any platform so far,  certain degree of control of the system is always needed, and this allows the implementation of the adiabatic paths to a certain precision.} As the other side of the story, the nontopological nature of our quantum computation protocols can also be exploited to realize a $T$ gate required for universal quantum computation \cite{magic, magic2, magic3, magic4, magic5}, which is otherwise impossible to construct via topologically protected braiding operations alone. To appreciate this point we can  skip steps 4-6 in the protocol described in Sec.~\ref{prot2}, leading to the net outcome $\gamma_\pi^L\rightarrow \gamma_\pi^L +\gamma_{0,2}^L$ and $\gamma_{0,2}^L\rightarrow  \gamma_{0,2}^L-\gamma_\pi^L$. This outcome is equivalent to the unitary $T_L=\exp\left[(\pi/8) \gamma_\pi^L \gamma_{0,2}^L\right]$, i.e., the $T$ gate acting on the first qubit. Similar approach can also be applied to realize $T_R$, the $T$ gate acting on the second qubit. {  Finally, it is noted that unlike other nontopological proposals for realizing $T$ gate \cite{magic3,magic4,magic5}, which are based on dynamical effect, our proposal is geometrical in nature and is thus expected to be more robust.}

Aside from examining the robustness of our scheme versus TQC, it is also important to point out that the novelty of our quantum computation scheme lies in the use of edge modes. Because our qubits are made of edge states, they do possess topological protection against some variations in the system parameters. This important advantage renders perfect fine tuning unnecessary and thus in principle provides advantages over other holonomic quantum computation proposals that do not rely on topological phases at all \cite{holo1,holo2,holo3,holo4,holo5,holo6,holo7}.

\section{Conclusion}
\label{conc}

 {  This work aims to advocate an alternative avenue of quantum computation by use of symmetry-protected edge modes of topological matter.}  A periodically driven quantum wire may host many zero and $\pi$ edge modes \cite{Derek1,Longwen1,Longwen2,kk3} being either as Majorana or fermionic excitations.
 Their dynamical phase contributions are trivial and hence adiabatic manipulations of these multiple edge modes associated with Floquet topological matter can be used for quantum information processing. %This is an exciting possibility not foreseen before.
  As the first step along this avenue,
 we exploit the coexistence of three pairs of Majorana edge modes in one single periodically driven quantum wire, equivalent to obtaining two local fermions and one nonlocal fermions as topologically protected edge modes.   The three pairs of Majorana edge modes can be used to encode two logical qubits, protected by both particle-hole and chiral symmetries.  Adiabatic protocols are designed to simulate the braiding between various pairs of Majorana modes, which then realizes several gate operations.  A means to readout these qubits is also proposed through introducing chiral-symmetry breaking terms into the system.  As an encouraging side result, we have also shown that our system can be scaled up, at least by adding more parallel quantum wires.  This then allows the implementation of entangling quantum gates.  To demonstrate the application of our quantum computation schemes, we have also constructed a quantum circuit to implement two simple quantum algorithms, which requires much less hardware resources as compared with previous work.   We have also briefly discussed potential realizations of our proposal in experiments. Understanding that there can be experimental challenges ahead but not yet identified,  we do not claim that any experimental realizations of this theoretical work would be straightforward at this point. {  However, the general features of our proposal, namely, the coexistence of different SPT phases and qubit encoding and manipulations, are hoped to motivate future studies on simpler systems that are easier to experimentally implement.} Finally, a comparison between our approach with that of TQC is also made.

 This paper indicates a possible new paradigm for realizing many logical qubits with minimal amount of physical resources on the hardware level.  Such kind of possibility, even still on the theoretical level,  is always stimulating towards the realization of a scalable quantum computer. As another consideration to scale up our quantum computation protocols, we call for future studies to explore the feasibility of using one single quantum wire to host and individually address more than two logical qubits.  A good starting point to achieve this is to consider systems capable of hosting many Majorana zero and $\pi$ modes, such as that considered in Ref.~\cite{kk3}.  More follow-up studies to that end will certainly enhance the marriage of two timely research topics as of today, namely, quantum computation and Floquet topological matter.  Indeed, this work should also serve as the first step to extend the idea of TQC to periodically driven systems. Following our discussion in Sec.~\ref{compare}, a possible future study is to devise computation protocols that can braid Majorana zero and $\pi$ modes purely through explicit monodromy, so as to unleash the full topological protection offered by braiding operations. It is expected that the combination of scalability of our proposal and the fault-tolerance nature of TQC approach may eventually lead to a full-fledged quantum computer based on topological edge modes.

\vspace{0.3cm}

\n {\bf Acknowledgements:}  J.G. acknowledges support by Singapore Ministry of Education Academic Research Fund Tier I (WBS No.~R-144-000-353-112) and by the Singapore NRF grant No.~NRF-NRFI2017-04 (WBS No.~R-144-000-378-281).

\appendix

\section{Derivation of Floquet non-Abelian Berry phase} \label{app1}

Following the notation in Sec.~\ref{holonomy}, we consider the application of $\mathcal{U}[\lambda(s)]$ on $|\varepsilon_n [\lambda(s-1)]\rangle$ as

\begin{equation}
\mathcal{U}[\lambda(s)] |\varepsilon_n [\lambda(s-1)]\rangle = V_n(\lambda) |\varepsilon_n [\lambda(s)]\rangle \;,
\label{evol}
\end{equation}

\noindent where we have used the fact that $\mathcal{U}[\lambda(s)]$ serves as a one-period propagator, combined with the adiabaticity condition that the state remains in a Floquet eigenstate associated with quasienergy $\varepsilon_n[\lambda(s)]$. $V_n(\lambda)\equiv \exp\left(\mathrm{i}\Omega_n(\lambda)\right)$ is a $k_n\times k_n$ path dependent unitary matrix which potentially rotates $|\varepsilon_n [\lambda(s)]\rangle$ in the degenerate subspace, thus generalizing the appearance of a global phase in the nondegenerate (Abelian) case.

Next, we expand

\begin{equation}
|\varepsilon_n [\lambda(s)]\rangle =\sum_{m} \exp\left(-\mathrm{i} \varepsilon_m (\lambda)T \right) V_m^\dagger(\lambda) C_m(\lambda) |\varepsilon_n [\lambda(s-1)]\rangle \;,
\label{expand}
\end{equation}

\noindent where $C_m(\lambda)$ is another $k_n\times k_n$ matrix that generalizes the spectral coefficients in the nondegenerate case. The left hand side of Eq.~(\ref{evol}) can be rewritten as

\begin{widetext}
	
\begin{eqnarray}
\mathcal{U}[\lambda(s)] |\varepsilon_n [\lambda(s-1)]\rangle &=& \mathcal{U}[\lambda(s)]  \mathcal{U}[\lambda(s-1)]^\dagger \mathcal{U}[\lambda(s-1)]|\varepsilon_n [\lambda(s-1)]\rangle \nonumber \\
&=& \exp\left(-\mathrm{i} \varepsilon_n[\lambda(s-1)] T \right) \mathcal{U}[\lambda(s)]  \mathcal{U}[\lambda(s-1)]^\dagger  |\varepsilon_n [\lambda(s-1)]\rangle \nonumber \\
&\approx& \exp\left(-\mathrm{i} \varepsilon_n[\lambda(s-1)]T \right)\left(\mathcal{I} +\frac{d\mathcal{U}}{d\lambda}\mathcal{U}^\dagger d\lambda\right)  |\varepsilon_n [\lambda(s-1)]\rangle \nonumber \\
&\approx & \exp\left(-\mathrm{i} \varepsilon_n[\lambda(s-1)]T\right)\exp\left(\frac{d\mathcal{U}}{d\lambda}\mathcal{U}^\dagger d\lambda\right) |\varepsilon_n [\lambda(s-1)]\rangle \;,
\label{simplify}
\end{eqnarray}

\end{widetext}

\noindent We can then combine Eqs.~(\ref{expand}) and (\ref{simplify}) with Eq.~(\ref{evol}), and apply both sides with $\langle \varepsilon_n [\lambda(s-1)]|$ from the left to obtain

\begin{equation}
\langle \varepsilon_n [\lambda(s-1)]| \exp\left(\frac{d\mathcal{U}}{d\lambda}\mathcal{U}^\dagger d\lambda\right) |\varepsilon_n [\lambda(s-1)]\rangle = C_n \;,
\end{equation}

\n where we have used the fact that matrix $C_n$ is only nondiagonal within a degenerate subspace, so that $\langle \varepsilon_n [\lambda(s-1)]| C_m |\varepsilon_m [\lambda(s-1)]\rangle=0$ if $m\neq n$. By spectral decomposing $\frac{d\mathcal{U}}{d\lambda}\mathcal{U}^\dagger$ and explicitly expanding the column vector defined in Sec.~\ref{holonomy}, we can derive the matrix coefficient of $C_n$ as

\begin{equation}
C_{n,\alpha\beta}=\exp\left(\langle \varepsilon_{n,\alpha} | \frac{d}{d\lambda} |\varepsilon_{n,\beta} \rangle d\lambda \right) \;.
\label{matrix}
\end{equation}

Finally, by recursively combining Eqs.~(\ref{expand}) and (\ref{matrix}), we arrive at

\begin{equation}
|\varepsilon_n (\lambda_\tau)\rangle =\mathcal{P} \exp\left(-\mathrm{i}\int_{\lambda_0}^{\lambda_\tau} \left[\mathcal{A}_n+\Omega_n+\varepsilon_n T \right] d\lambda \right) |\varepsilon_n (\lambda_0)\rangle\;,
\end{equation}

\n where $\mathcal{P}$ is the path ordering operator, and $\mathcal{A}_{n,\alpha,\beta}=\mathrm{i} \langle \varepsilon_{n,\alpha} | \frac{d}{d\lambda} |\varepsilon_{n,\beta} \rangle$ is the non-Abelian Berry connection.

\section{Evolution of Majorana modes during $\gamma_{0,1}^L$ and $\gamma_{0,2}^L$ braiding protocol} \label{app2}

For all the steps presented in Sec.~\ref{prot1}, we are able to analytically keep track the evolution of all Majorana modes by recursively solving $[\mathcal{U}^{(S)}(\phi),\gamma_0]=0$ and $\left\lbrace \mathcal{U}^{(S)},\gamma_\pi \right\rbrace=0$ for Majorana zero and $\pi$ modes respectively, where $\mathcal{U}^{(S)}(\phi)$ is the Floquet operator at step $S=1,\cdots, 6$, which can be written as
{ 
	
\begin{equation}
\mathcal{U}^{(S)}(\phi)=\mathcal{U}_{H_2}^{(S)}(\phi)\times \mathcal{U}_{H_1}^{(S)}(\phi) \;,
\label{fstep}
\end{equation}
	
\noindent where $\mathcal{U}_{H_2}^{(S)}(\phi)=\exp\left(-\mathrm{i} H_{2}^{(S)}(\phi) T/2 \right)$, $\mathcal{U}_{H_1}^{(S)}(\phi)=\exp\left(-\mathrm{i} H_1^{(S)}(\phi) T/2\right)$, $H_1^{(S)}(\phi)$ and $H_2^{(S)}(\phi)$ are the deformation of the two Hamiltonian in Eq.~(\ref{H}) when subjected to the adiabatic modulation in hopping and pairing strength as prescribed in Sec.~\ref{prot1}.} It is convenient to express $H_1^{(S)}$ and $H_2^{(S)}$ in Majorana basis as (keeping only terms in the first two lattice sites for brevity)

\begin{widetext}
\begin{eqnarray}
H_1^{(1)} T&=&  \mathrm{i} \frac{\pi}{2} \left[  \gamma_{B,1;A,1}^{\alpha;\beta}+ \gamma_{A,2;B,1}^{\alpha;\beta} + \Gamma_1 \right] \;,\nonumber \\
H_1^{(2)} T &=& \mathrm{i} \frac{\pi}{2} \left[c(\phi)  \gamma_{A,2;B,1}^{\alpha;\beta}+s(\phi)  \gamma_{B,1;A,1}^{\beta;\alpha} + \gamma_{B,1;A,1}^{\alpha;\beta}+\Gamma_1 \right] \;,\nonumber \\
%&& \left. +\gamma_{B,2;A,2}^{\alpha;\beta}+\gamma_{A,3;B,2}^{\alpha;\beta} \right] \;, \nonumber \\
H_1^{(3)} T&=& H_1^{(2)} T\;, \nonumber \\
H_1^{(4)} T &=& \mathrm{i} \frac{\pi}{2} \left[ c(\phi) \gamma_{B,1;A,1}^{\beta;\alpha}+ \gamma_{B,1;A,1}^{\alpha;\beta}  + \Gamma_1 \right] \;, \nonumber \\
\end{eqnarray}

\begin{eqnarray}
H_1^{(5)} T &=& \mathrm{i} \frac{\pi}{2}\left[s(\phi) \gamma_{A,2;B,1}^{\alpha;\beta}+\gamma_{B,1;A,1}^{\alpha;\beta}+\Gamma_1\right] \;, \nonumber \\
H_1^{(6)} T &=& \mathrm{i} \frac{\pi}{2} \left[\gamma_{A,2;B,1}^{\alpha;\beta}+\gamma_{B,1;A,1}^{\alpha;\beta}+\Gamma_1 \right] \;, \nonumber \\
H_2^{(1)} T &=& \mathrm{i} \pi \left[c(\phi) \gamma_{B,1;A,2}^{\beta;\beta}+s(\phi)\gamma_{A,1;B,1}^{\beta;\beta}+\gamma_{B,1;A,2}^{\alpha;\alpha}+\Gamma_2 \right]\;, \nonumber \\
%&& \left.+ \gamma_{B,2;A,3}^{\alpha;\alpha}+\gamma_{B,2;A,3}^{\beta;\beta} \right] \;, \nonumber \\
H_2^{(2)} T &=& \mathrm{i} \pi \left[s(\phi) \gamma_{A,1;B,1}^{\alpha;\alpha}+c(\phi) \gamma_{B,1;A,2}^{\alpha;\alpha}+ \gamma_{A,1;B,1}^{\beta;\beta} +\Gamma_2 \right] \;,\nonumber \\
%&& \left.+ \gamma_{B,2;A,3}^{\alpha;\alpha}+\gamma_{B,2;A,3}^{\beta;\beta}\right] \;, \nonumber \\
H_2^{(3)} T &=& \mathrm{i} \pi \left[ c(\phi)\gamma_{A,1;B,1}^{\alpha;\alpha}+s(\phi) \gamma_{B,1;A,2}^{\alpha;\beta}+\gamma_{A,1;B,1}^{\beta;\beta}+\Gamma_2 \right]\; \nonumber \\
%&& \left. +\gamma_{B,2;A,3}^{\beta;\beta} + \gamma_{B,2;A,3}^{\alpha;\alpha} \right] \;,\nonumber \\
H_2^{(4)} T &=& \mathrm{i} \pi \left[\gamma_{A,1;B,1}^{\beta;\beta}+ \gamma_{B,1;A,2}^{\alpha;\beta}+\Gamma_2 \right] \;, \nonumber \\
H_2^{(5)} T &=& \mathrm{i}\pi \left[s(\phi)\gamma_{B,1;A,2}^{\alpha;\alpha}+c(\phi)\gamma_{B,1;A,2}^{\alpha;\beta} +\gamma_{A,1;B,1}^{\beta;\beta}+\Gamma_2 \right]\;, \nonumber \\
%&& \left. \gamma_{B,2;A,3}^{\beta;\beta}+\gamma_{B,2;A,3}^{\alpha;\alpha}\right] \;, \nonumber \\
H_2^{(6)} T &=& \mathrm{i} \pi \left[s(\phi) \gamma_{B,1;A,2}^{\beta;\beta}+c(\phi) \gamma_{A,1;B,1}^{\beta;\beta}+ \gamma_{B,1;A,2}^{\alpha;\alpha}+\Gamma_2 \right] \;, \nonumber \\
%&& \left. \gamma_{B,2;A,3}^{\beta;\beta}+\gamma_{B,2;A,3}^{\alpha;\alpha}\right] \;,
\end{eqnarray}
\end{widetext}

\noindent where $\Gamma_1 =\gamma_{B,2;A,2}^{\alpha;\beta}+\gamma_{A,3;B,2}^{\alpha;\beta}$, $\Gamma_2=\gamma_{B,2;A,3}^{\alpha;\alpha}+\gamma_{B,2;A,3}^{\beta;\beta}$, $\gamma_{C,j;D,k}^{\mu;\nu}$ stands for $\gamma_{C,j}^\mu \gamma_{D,k}^\nu$, $C,D\in\left\lbrace A,B\right\rbrace$, $\mu,\nu\in \left\lbrace\alpha,\beta\right\rbrace$, $j,k\in \left\lbrace 1, 2, 3\right\rbrace$ denote the lattice site, $s(\phi)$ and $c(\phi)$ stand for $\sin(\phi)$ and $\cos(\phi)$ respectively.

Rather than showing the full derivation of the Majorana modes from the recurrence relation, we will instead show the form of the Majorana modes affected by the deformation at each step, and briefly verify them by commuting with $\mathcal{U}^{(S)}(\phi)$. The latter can be done analytically by using the following two facts.

{ 
\begin{enumerate}
\item Most of the terms in $H_1^{(S)}$ and $H_2^{(S)}$ commute with one another. This allows us to write Eq.~(\ref{fstep}) as products of many exponentials. For example, given a Hamiltonian of the form $H=\mathrm{i} \left[\gamma_1\gamma_2+\gamma_3 \gamma_4\right]$, the associated Floquet operator can be written as

\begin{equation*}
\mathcal{U}=\exp\left(-\mathrm{i} T\gamma_1\gamma_2 \right)\times \exp\left(-\mathrm{i} T\gamma_3\gamma_4 \right)\;.
\end{equation*}
\item The application of each exponential on a given Majorana operator $\gamma_1$ satisfies

\begin{eqnarray*}
\exp\left(\theta \gamma_1 \gamma_2\right) \gamma_1 \exp\left(-\theta \gamma_1 \gamma_2\right) &=& \cos(2\theta)\gamma_1 -\sin(2\theta) \gamma_2 \;,\\
\exp\left(\theta \gamma_2 \gamma_3\right) \gamma_1 \exp\left(-\theta \gamma_2 \gamma_3\right) &=& \gamma_1\;,
\end{eqnarray*}
\end{enumerate}

\n which can be proven using the identity Eq.~(\ref{id}).

\textit{Step 1:}

\begin{eqnarray}
%\gamma_\pi^L(\phi) &=& \frac{1}{\sqrt{2}}\left[\left(\gamma_{A,1}^\beta-\gamma_{B,1}^\alpha\right)\cos\phi + \left(\gamma_{A,2}^\beta-\gamma_{B,2}^\alpha\right)\sin\phi\right] \;,\nonumber \\
\gamma_\pi^L(\phi) &=&\left[c(\phi)\gamma_{A,1}^\beta+s(\phi)\gamma_{A,2}^\beta\right] + \left[c(\phi)\gamma_{B,1}^\alpha+s(\phi)\gamma_{B,2}^\alpha\right] \;,\nonumber \\
%\gamma_{0,2}^L(\phi) &=& \frac{1}{\sqrt{2}}\left[\left(\gamma_{A,1}^\beta+\gamma_{B,1}^\alpha\right)\cos\phi + \left(\gamma_{A,2}^\beta+\gamma_{B,2}^\alpha\right)\sin\phi\right]\;, \nonumber \\
\gamma_{0,2}^L(\phi) &=&\left[c(\phi)\gamma_{A,1}^\beta+s(\phi)\gamma_{A,2}^\beta\right] - \left[c(\phi)\gamma_{B,1}^\alpha+s(\phi)\gamma_{B,2}^\alpha\right] \;,\nonumber \\
\end{eqnarray}

\n where we have suppressed the normalization factor for brevity here and for the rest of the steps. In particular, two Majorana operators are involved in this step, which are $\gamma_1 =c(\phi)\gamma_{A,1}^\beta+s(\phi)\gamma_{A,2}^\beta$ and $\gamma_2=c(\phi)\gamma_{B,1}^\alpha+s(\phi)\gamma_{B,2}^\alpha $. The application of $\mathcal{U}^{(1)}(\phi)$ to these Majorana operators can be written as (using the two facts above),

\begin{widetext}

\begin{eqnarray*}
\mathcal{U}^{(1)}(\phi)^\dagger \gamma_1 \mathcal{U}^{(1)}(\phi) &=& \mathcal{U}_{H_1}^{(1)}(\phi)^\dagger \mathcal{U}_{H_2}^{(1)}(\phi)^\dagger \gamma_1 \mathcal{U}_{H_2}^{(1)}(\phi) \mathcal{U}_{H_1}^{(1)}(\phi) \\
&=& \mathcal{U}_{H_1}^{(1)}(\phi)^\dagger \gamma_1 \mathcal{U}_{H_1}^{(1)}(\phi) \\
&=& c(\phi) \exp\left(- \pi/4 \gamma_{B,1;A,1}^{\alpha;\beta}\right) \gamma_{A,1}^\beta \exp\left(\pi/4 \gamma_{B,1;A,1}^{\alpha;\beta}\right) +s(\phi) \exp\left(- \pi/4 \gamma_{B,2;A,2}^{\alpha;\beta}\right) \gamma_{A,2}^\beta \exp\left(\pi/4 \gamma_{B,2;A,2}^{\alpha;\beta}\right)\\
&=& -c(\phi) \gamma_{B,1}^\alpha-s(\phi) \gamma_{B,2}^\alpha \\
&=&-\gamma_2 \;, \\
\end{eqnarray*}

\begin{eqnarray*}
\mathcal{U}^{(1)}(\phi)^\dagger \gamma_2 \mathcal{U}^{(1)}(\phi) &=& \mathcal{U}_{H_1}^{(1)}(\phi)^\dagger \mathcal{U}_{H_2}^{(1)}(\phi)^\dagger \gamma_2 \mathcal{U}_{H_2}^{(1)}(\phi) \mathcal{U}_{H_1}^{(1)}(\phi) \\
&=& \mathcal{U}_{H_1}^{(1)}(\phi)^\dagger \left[c(\phi) \exp\left(- \pi/2 \gamma_{B,1;A,2}^{\alpha;\alpha}\right)\gamma_{B,1}^\alpha \exp\left( \pi/2 \gamma_{B,1;A,2}^{\alpha;\alpha}\right) \right. \\
&& \left.+s(\phi) \exp\left(- \pi/2 \gamma_{B,2;A,3}^{\alpha;\alpha}\right)\gamma_{B,2}^\alpha \exp\left( \pi/2 \gamma_{B,2;A,3}^{\alpha;\alpha}\right)\right] \mathcal{U}_{H_1}^{(1)}(\phi) \\
&=& -  \mathcal{U}_{H_1}^{(1)}(\phi)^\dagger \left[c(\phi)\gamma_{B,1}^\alpha+s(\phi)\gamma_{B,2}^\alpha\right] \mathcal{U}_{H_1}^{(1)}(\phi)^\dagger \\
&=& -c(\phi) \exp\left(- \pi/4 \gamma_{B,1;A,1}^{\alpha;\beta}\right) \gamma_{B,1}^\alpha \exp\left(\pi/4 \gamma_{B,1;A,1}^{\alpha;\beta}\right) -s(\phi) \exp\left(- \pi/4 \gamma_{B,2;A,2}^{\alpha;\beta}\right) \gamma_{B,2}^\alpha \exp\left(\pi/4 \gamma_{B,2;A,2}^{\alpha;\beta}\right)\\
&=& -\gamma_1 \;.
\end{eqnarray*}

\end{widetext}

%This can be verified by first noting, by using the two facts above, that $\exp\left(-\mathrm{i} H_1^{(2)} T/2\right) $ interchanges $(\gamma_{A,1}^\beta,\gamma_{A,2}^\beta)\rightarrow -(\gamma_{B,1}^\alpha,\gamma_{B,2}^\alpha)$ and $(\gamma_{B,1}^\alpha,\gamma_{B,2}^\alpha)\rightarrow (\gamma_{A,1}^\beta,\gamma_{A,2}^\beta)$. On the other hand, $\exp\left(-\mathrm{i} H_2^{(2)} T/2\right)$ flips the sign of $\gamma_{B,1}^\alpha$ and $\gamma_{B,2}^\alpha$. Together, they yield the correct transformation $\mathcal{U}^{(2)\dagger} \gamma_\pi^L\mathcal{U}^{(2)} = -\gamma_\pi^L$ and $\mathcal{U}^{(2)\dagger}\gamma_{0,2}^L\mathcal{U}^{(2)} = \gamma_{0,2}^L$.

\n Symmetric and antisymmetric superpositions of $\gamma_1$ and $\gamma_2$ thus anticommute and commute with $\mathcal{U}^{(1)}(\phi)$ and correspond to Majorana $\pi$ and zero modes respectively, as claimed above.
}

\textit{Step 2:} $\gamma_{0,1}^L(\phi)=s(\phi) \gamma_{A,2}^\alpha+ c(\phi) \gamma_{A,1}^\alpha$. This is easily verified by noting that it commutes with both $H_1^{(2)}$ and $H_2^{(2)}${,   thereby with $\mathcal{U}^{(2)}(\phi)$ too.

\textit{Step 3:}

\begin{eqnarray}
%\gamma_\pi^L(\phi) &=& \frac{1}{\sqrt{2}}\left[\left(\gamma_{A,2}^\beta-\gamma_{B,2}^\alpha\right)\cos\phi + \left(\gamma_{A,1}^\alpha-\gamma_{B,1}^\beta\right)\sin\phi\right] \;,\nonumber \\
\gamma_\pi^L(\phi) &=&\left[c(\phi)\gamma_{A,2}^\beta+s(\phi)\gamma_{A,1}^\alpha\right] + \left[c(\phi)\gamma_{B,2}^\alpha+s(\phi)\gamma_{B,1}^\beta\right] \;,\nonumber \\
%\gamma_{0,2}^L(\phi) &=& \frac{1}{\sqrt{2}}\left[\left(\gamma_{A,2}^\beta+\gamma_{B,2}^\alpha\right)\cos\phi + \left(\gamma_{A,1}^\alpha+\gamma_{B,1}^\beta\right)\sin\phi\right]\;, \nonumber \\
\gamma_{0,2}^L(\phi) &=&\left[c(\phi)\gamma_{A,2}^\beta+s(\phi)\gamma_{A,1}^\alpha\right] - \left[c(\phi)\gamma_{B,2}^\alpha+s(\phi)\gamma_{B,1}^\beta\right] \;.\nonumber \\
\end{eqnarray}

\n This can be verified in the same way as in step 1. That is, we first denote $\gamma_1=c(\phi)\gamma_{A,2}^\beta+s(\phi)\gamma_{A,1}^\alpha$ and $\gamma_2 = c(\phi)\gamma_{B,2}^\alpha+s(\phi)\gamma_{B,1}^\beta$ respectively. We can then verify the application of $\mathcal{U}^{(3)}(\phi)$ on $\gamma_1$ and $\gamma_2$ as%That is, $\exp\left(-\mathrm{i} H_1^{(3)} T/2\right) $ interchanges $(\gamma_{A,2}^\beta,\gamma_{A,1}^\alpha)\rightarrow -(\gamma_{B,2}^\alpha,\gamma_{B,1}^\beta)$ and $(\gamma_{B,2}^\alpha,\gamma_{B,1}^\beta)\rightarrow (\gamma_{A,2}^\beta,\gamma_{A,1}^\alpha)$, $\exp\left(-\mathrm{i} H_2^{(3)} T/2\right) $ flips the sign of $\gamma_{B,2}^\alpha$ and $\gamma_{B,1}^\beta$, resulting in $\mathcal{U}^{(3)\dagger} \gamma_\pi^L\mathcal{U}^{(3)} = -\gamma_\pi^L$ and $\mathcal{U}^{(3)\dagger}\gamma_{0,2}^L\mathcal{U}^{(3)} = \gamma_{0,2}^L$.

\begin{widetext}
	
	\begin{eqnarray*}
		\mathcal{U}^{(3)}(\phi)^\dagger \gamma_1 \mathcal{U}^{(3)}(\phi) &=& c(\phi) \exp\left(- \pi/4 \gamma_{B,2;A,2}^{\alpha;\beta}\right) \gamma_{A,2}^\beta \exp\left(\pi/4 \gamma_{B,2;A,2}^{\alpha;\beta}\right) +s(\phi) \exp\left(- \pi/4 \gamma_{B,1;A,1}^{\beta;\alpha}\right) \gamma_{A,1}^\alpha \exp\left(\pi/4 \gamma_{B,1;A,1}^{\beta;\alpha}\right)\\
		&=& -c(\phi) \gamma_{B,2}^\alpha-s(\phi) \gamma_{B,1}^\beta \\
		&=&-\gamma_2 \;, \\
		\mathcal{U}^{(3)}(\phi)^\dagger \gamma_2 \mathcal{U}^{(3)}(\phi) &=& \mathcal{U}_{H_1}^{(3)}(\phi)^\dagger \left[c(\phi) \exp\left(- \pi/2 \gamma_{B,2;A,3}^{\alpha;\alpha}\right)\gamma_{B,2}^\alpha \exp\left( \pi/2 \gamma_{B,2;A,3}^{\alpha;\alpha}\right) \right. \\
		&& \left.+s(\phi) \exp\left(- \pi/2 \gamma_{A,1;B,1}^{\beta;\beta}\right)\gamma_{B,1}^\beta \exp\left( \pi/2 \gamma_{A,1;B,1}^{\beta;\beta}\right)\right] \mathcal{U}_{H_1}^{(3)}(\phi) \\
		&=& -  \mathcal{U}_{H_1}^{(3)}(\phi)^\dagger \left[c(\phi)\gamma_{B,2}^\alpha+s(\phi)\gamma_{B,1}^\beta\right] \mathcal{U}_{H_1}^{(3)}(\phi)^\dagger \\
		&=& -c(\phi) \exp\left(- \pi/4 \gamma_{B,2;A,2}^{\alpha;\beta}\right) \gamma_{B,2}^\alpha \exp\left(\pi/4 \gamma_{B,2;A,2}^{\alpha;\beta}\right) -s(\phi) \exp\left(- \pi/4 \gamma_{B,1;A,1}^{\beta;\alpha}\right) \gamma_{B,1}^\beta \exp\left(\pi/4 \gamma_{B,1;A,1}^{\beta;\alpha}\right)\\
		&=& -\gamma_1 \;,
	\end{eqnarray*}
	
\end{widetext}

\n Similar to step 1, symmetric and antisymmetric superpositions of $\gamma_1$ and $\gamma_2$ (i.e., $\gamma_\pi^L(\phi)$ and $\gamma_{0,2}^L(\phi)$) thus anticommute and commute with $\mathcal{U}^{(3)}(\phi)$

\textit{Step 4:}

\begin{eqnarray}
\gamma_\pi^L (\phi)&=& s\left(\frac{\pi}{4} c(\phi) \right) \gamma_{A,1}^\alpha + c\left(\frac{\pi}{4}c(\phi) \right) \gamma_{B,1}^\beta \;, \nonumber \\
\gamma_{0,2}^L (\phi)&=& c\left(\frac{\pi}{4}c(\phi) \right) \gamma_{A,1}^\alpha - s\left(\frac{\pi}{4}c(\phi) \right) \gamma_{B,1}^\beta \;.
\end{eqnarray}

These can be verified by applying $\mathcal{U}^{(4)}(\phi)$ directly to $\gamma_\pi^L(\phi)$ and $\gamma_{0,2}^L$,

\begin{widetext}
	
\begin{eqnarray*}
\mathcal{U}^{(4)}(\phi)^\dagger \gamma_\pi^L \mathcal{U}^{(4)}(\phi) &=& \mathcal{U}_{H_1}^{(4)}(\phi)^\dagger \left[s\left(\frac{\pi}{4} c(\phi)\right) \gamma_{A,1}^\alpha +c\left(\frac{\pi}{4} c(\phi)\right) \exp\left(- \pi/2 \gamma_{A,1;B,1}^{\beta;\beta}\right) \gamma_{B,1}^\beta \exp\left(\pi/2 \gamma_{A,1;B,1}^{\beta;\beta}\right)\right] \mathcal{U}_{H_1}^{(4)}(\phi) \\
&=& \mathcal{U}_{H_1}^{(4)}(\phi)^\dagger \left[s\left(\frac{\pi}{4} c(\phi)\right) \gamma_{A,1}^\alpha -c\left(\frac{\pi}{4} c(\phi)\right)  \gamma_{B,1}^\beta \right] \mathcal{U}_{H_1}^{(4)}(\phi) \\
&=& \exp\left(- \pi/4 c(\phi) \gamma_{B,1;A,1}^{\beta;\alpha}\right) \left[s\left(\frac{\pi}{4} c(\phi)\right) \gamma_{A,1}^\alpha -c\left(\frac{\pi}{4} c(\phi)\right)  \gamma_{B,1}^\beta \right] \exp\left( \pi/4 c(\phi) \gamma_{B,1;A,1}^{\beta;\alpha}\right) \\
&=& -s\left(\frac{\pi}{4} c(\phi) \right) \gamma_{A,1}^\alpha - c\left(\frac{\pi}{4}c(\phi) \right) \gamma_{B,1}^\beta \\
&=& -\gamma_\pi^L \;, \\
\mathcal{U}^{(4)}(\phi)^\dagger \gamma_{0,2}^L \mathcal{U}^{(4)}(\phi) &=& \mathcal{U}_{H_1}^{(4)}(\phi)^\dagger \left[c\left(\frac{\pi}{4} c(\phi)\right) \gamma_{A,1}^\alpha -s\left(\frac{\pi}{4} c(\phi)\right) \exp\left(- \pi/2 \gamma_{A,1;B,1}^{\beta;\beta}\right) \gamma_{B,1}^\beta \exp\left(\pi/2 \gamma_{A,1;B,1}^{\beta;\beta}\right)\right] \mathcal{U}_{H_1}^{(4)}(\phi) \\
&=& \mathcal{U}_{H_1}^{(4)}(\phi)^\dagger \left[c\left(\frac{\pi}{4} c(\phi)\right) \gamma_{A,1}^\alpha +s\left(\frac{\pi}{4} c(\phi)\right)  \gamma_{B,1}^\beta \right] \mathcal{U}_{H_1}^{(4)}(\phi) \\
&=& \exp\left(- \pi/4 c(\phi) \gamma_{B,1;A,1}^{\beta;\alpha}\right) \left[c\left(\frac{\pi}{4} c(\phi)\right) \gamma_{A,1}^\alpha +s\left(\frac{\pi}{4} c(\phi)\right)  \gamma_{B,1}^\beta \right] \exp\left( \pi/4 c(\phi) \gamma_{B,1;A,1}^{\beta;\alpha}\right) \\
&=& c\left(\frac{\pi}{4} c(\phi) \right) \gamma_{A,1}^\alpha + s\left(\frac{\pi}{4}c(\phi) \right) \gamma_{B,1}^\beta \\
&=& -\gamma_{0,2}^L \;,
\end{eqnarray*}

\end{widetext}	
	
	\n where we have used Eq.~(\ref{id}) to arrive at the second last line of each expansion above.
%This can be verified by noting that $\exp\left(-\mathrm{i} H_1^{(4)} T/2\right) $ maps $\gamma_\pi^L\rightarrow -s\left(\frac{\pi}{4} c(\phi) \right) \gamma_{A,1}^\alpha - c\left(\frac{\pi}{4}c(\phi) \right) \gamma_{B,1}^\beta$ and $\gamma_{0,2}^L (\phi)\rightarrow c\left(\frac{\pi}{4}c(\phi) \right) \gamma_{A,1}^\alpha - s\left(\frac{\pi}{4}c(\phi) \right) \gamma_{B,1}^\beta$. On the other hand, $\exp\left(-\mathrm{i} H_2^{(4)} T/2\right) $ flips the sign of $\gamma_{B,1}^\beta$.

\textit{Step 5:}

\begin{eqnarray}
%\gamma_\pi^L (\phi)&=& -\left[\sin\left(\frac{\pi}{4}\sin\phi \right) \gamma_{A,2}^\alpha + \cos\left(\frac{\pi}{4}\sin\phi \right) \gamma_{B,1}^\beta\right]\cos\phi \nonumber \\
%&& +\left[\sin\left(\frac{\pi}{4}\sin\phi \right) \gamma_{A,2}^\beta - \cos\left(\frac{\pi}{4}\sin\phi \right) \gamma_{B,2}^\alpha\right]\sin\phi \;, \nonumber \\
\gamma_\pi^L(\phi) &=&\left[s(\phi)\left(\gamma_{A,2}^\beta+\gamma_{B,2}^\alpha\right)-c(\phi)\gamma_{A,2}^\alpha\right]s\left(\frac{\pi}{4}s(\phi)\right) \nonumber \\
&& + c(\phi)\gamma_{B,1}^\beta  c\left(\frac{\pi}{4}s(\phi)\right)\;,\nonumber \\
%\gamma_{0,1}^L (\phi)&=& \left[\cos\left(\frac{\pi}{4}\sin\phi \right) \gamma_{A,2}^\alpha - \sin\left(\frac{\pi}{4}\sin\phi \right) \gamma_{B,1}^\beta\right]\cos\phi \nonumber \\
%&& -\left[\cos\left(\frac{\pi}{4}\sin\phi \right) \gamma_{A,2}^\beta + \sin\left(\frac{\pi}{4}\sin\phi \right) \gamma_{B,2}^\alpha\right]\sin\phi \;, \nonumber \\
\gamma_{0,1}^L(\phi) &=&\left[c(\phi)\gamma_{A,2}^\alpha-s(\phi)\left(\gamma_{A,2}^\beta-\gamma_{B,2}^\alpha\right)\right]c\left(\frac{\pi}{4}s(\phi)\right) \nonumber \\
&& + c(\phi)\gamma_{B,1}^\beta s\left(\frac{\pi}{4}s(\phi)\right)\;.
\end{eqnarray}

\n By applying $\mathcal{U}^{(5)}(\phi)$ directly to $\gamma_\pi^L(\phi)$ and $\gamma_{0,1}^L$,

\begin{widetext}

\begin{eqnarray*}
\mathcal{U}^{(5)}(\phi)^\dagger \gamma_\pi^L(\phi) \mathcal{U}^{(5)}(\phi) &=& \mathcal{U}_{H_1}^{(5)}(\phi)^\dagger \left\lbrace \left[ s(\phi)\left(\gamma_{A,2}^\beta+\exp\left(- \pi/2 \gamma_{B,2;A,3}^{\alpha;\alpha}\right)\gamma_{B,2}^\alpha \exp\left( \pi/2  \gamma_{B,2;A,3}^{\alpha;\alpha}\right)\right)-c(\phi)\gamma_{A,2}^\alpha\right]s\left(\frac{\pi}{4}s(\phi)\right) \right. \\
&& \left. + c(\phi)\exp\left( -\pi/2  \gamma_{A,1;B,1}^{\beta;\beta}\right) \gamma_{B,1}^\beta  \exp\left( \pi/2  \gamma_{A,1;B,1}^{\beta;\beta}\right)c\left(\frac{\pi}{4}s(\phi)\right) \right\rbrace \mathcal{U}_{H_1}^{(5)}(\phi) \\
&=&  \mathcal{U}_{H_1}^{(5)}(\phi)^\dagger \left\lbrace \left[s(\phi)\left(\gamma_{A,2}^\beta-\gamma_{B,2}^\alpha \right)-c(\phi)\gamma_{A,2}^\alpha\right]s\left(\frac{\pi}{4}s(\phi)\right)- c(\phi)\gamma_{B,1}^\beta  c\left(\frac{\pi}{4}s(\phi)\right) \right\rbrace \mathcal{U}_{H_1}^{(5)}(\phi) \\
&=& s(\phi)s\left(\frac{\pi}{4}s(\phi)\right)\exp\left(- \pi/4 \gamma_{B,2;A,2}^{\alpha;\beta}\right) \left[\gamma_{A,2}^\beta-\gamma_{B,2}^\alpha\right] \exp\left(\pi/4  \gamma_{B,2;A,2}^{\alpha;\beta}\right) \\
&& -c(\phi) \exp\left(- \pi/4 s(\phi) \gamma_{A,2;B,1}^{\alpha;\beta}\right) \left[ c\left(\frac{\pi}{4}s(\phi)\right)\gamma_{B,1}^\beta  +s\left(\frac{\pi}{4}s(\phi)\right)\gamma_{A,2}^\alpha \right] \exp\left(\pi/4 s(\phi) \gamma_{A,2;B,1}^{\alpha;\beta}\right) \\
&=& -\left[s(\phi) \left(\gamma_{A,2}^\beta+\gamma_{B,2}^\alpha\right)-c(\phi)\gamma_{A,2}^\alpha\right]s\left(\frac{\pi}{4}s(\phi)\right)-c(\phi)\gamma_{B,1}^\beta  c\left(\frac{\pi}{4}s(\phi)\right) \\
&=& -\gamma_\pi^L \;,\\
\mathcal{U}^{(5)}(\phi)^\dagger \gamma_{0,1}^L(\phi) \mathcal{U}^{(5)}(\phi)&=& \mathcal{U}_{H_1}^{(5)}(\phi)^\dagger \left\lbrace \left[c(\phi)\gamma_{A,2}^\alpha- s(\phi)\left(\gamma_{A,2}^\beta-\exp\left(- \pi/2 \gamma_{B,2;A,3}^{\alpha;\alpha}\right)\gamma_{B,2}^\alpha \exp\left( \pi/2  \gamma_{B,2;A,3}^{\alpha;\alpha}\right)\right)\right]c\left(\frac{\pi}{4}s(\phi)\right) \right. \\
&& \left. +c(\phi)\exp\left( -\pi/2  \gamma_{A,1;B,1}^{\beta;\beta}\right) \gamma_{B,1}^\beta  \exp\left( \pi/2  \gamma_{A,1;B,1}^{\beta;\beta}\right)s\left(\frac{\pi}{4}s(\phi)\right) \right\rbrace \mathcal{U}_{H_1}^{(5)}(\phi) \\
&=& \mathcal{U}_{H_1}^{(5)}(\phi)^\dagger \left\lbrace \left[c(\phi)\gamma_{A,2}^\alpha-s(\phi)\left(\gamma_{A,2}^\beta+\gamma_{B,2}^\alpha \right)\right]c\left(\frac{\pi}{4}s(\phi)\right)- c(\phi)\gamma_{B,1}^\beta  s\left(\frac{\pi}{4}s(\phi)\right) \right\rbrace \mathcal{U}_{H_1}^{(5)}(\phi) \\
&=& -s(\phi)c\left(\frac{\pi}{4}s(\phi)\right)\exp\left(- \pi/4 \gamma_{B,2;A,2}^{\alpha;\beta}\right) \left[\gamma_{A,2}^\beta+\gamma_{B,2}^\alpha\right] \exp\left(\pi/4  \gamma_{B,2;A,2}^{\alpha;\beta}\right) \\
&& -c(\phi) \exp\left(- \pi/4 s(\phi) \gamma_{A,2;B,1}^{\alpha;\beta}\right) \left[ s\left(\frac{\pi}{4}s(\phi)\right)\gamma_{B,1}^\beta  -c\left(\frac{\pi}{4}s(\phi)\right)\gamma_{A,2}^\alpha \right] \exp\left(\pi/4 s(\phi) \gamma_{A,2;B,1}^{\alpha;\beta}\right) \\
&=& \gamma_{0,1}^L \;.
\end{eqnarray*}
	
\end{widetext}

 %This can be verified by noting that $\exp\left(-\mathrm{i} H_1^{(5)} T/2\right) $ maps

%\begin{eqnarray}
%\gamma_\pi^L(\phi) &\rightarrow &\left[-s(\phi)\left(\gamma_{A,2}^\beta+\gamma_{B,2}^\alpha\right)+c(\phi)\gamma_{A,2}^\alpha\right]s\left(\frac{\pi}{4}s(\phi)\right) \nonumber \\
%&& + c(\phi)\gamma_{B,1}^\beta  c\left(\frac{\pi}{4}s(\phi)\right)\;,\nonumber \\
%\gamma_{0,1}^L(\phi) %&\rightarrow&\left[c(\phi)\gamma_{A,2}^\alpha-s(\phi)\left(\gamma_{A,2}^\beta-\gamma_{B,2}^\alpha\right)\right]c\left(\frac{\pi}{4}s(\phi)\right) \nonumber \\
%&& - c(\phi)\gamma_{B,1}^\beta s\left(\frac{\pi}{4}s(\phi)\right)\;.
%\end{eqnarray}

%\n On the other hand, $\exp\left(-\mathrm{i} H_2^{(5)} T/2\right) $ flips the sign of $\gamma_{B,2}^\alpha$ and $\gamma_{B,1}^\beta$.

\textit{Step 6:}

\begin{eqnarray}
%\gamma_\pi^L (\phi) &=& \frac{1}{\sqrt{2}} \left[\left(\gamma_{A,2}^\beta-\gamma_{B,2}^\alpha\right)\cos\phi+\left(\gamma_{A,1}^\beta-\gamma_{B,1}^\alpha\right)\sin\phi\right]\;,\nonumber \\
\gamma_\pi^L(\phi) &=&\left[c(\phi)\gamma_{A,2}^\beta+s(\phi)\gamma_{A,1}^\beta\right] + \left[c(\phi)\gamma_{B,2}^\alpha+s(\phi)\gamma_{B,1}^\alpha\right] \;,\nonumber \\
%\gamma_{0,1}^L (\phi) &=& \frac{1}{\sqrt{2}} \left[\left(\gamma_{A,2}^\beta+\gamma_{B,2}^\alpha\right)\cos\phi+\left(\gamma_{A,1}^\beta+\gamma_{B,1}^\alpha\right)\sin\phi\right]\;.\nonumber \\
\gamma_{0,1}^L(\phi) &=&\left[c(\phi)\gamma_{A,2}^\beta+s(\phi)\gamma_{A,1}^\beta\right] - \left[c(\phi)\gamma_{B,2}^\alpha+s(\phi)\gamma_{B,1}^\alpha\right] \;.\nonumber \\
\end{eqnarray}

\n Following steps 1 and 3, define $\gamma_1=c(\phi)\gamma_{A,2}^\beta+s(\phi)\gamma_{A,1}^\beta$ and $\gamma_2=c(\phi)\gamma_{B,2}^\alpha+s(\phi)\gamma_{B,1}^\alpha$. It follows that

\begin{widetext}

\begin{eqnarray*}
\mathcal{U}^{(6)}(\phi)^\dagger \gamma_1 \mathcal{U}^{(6)}(\phi) &=& c(\phi) \exp\left(- \pi/4 \gamma_{B,2;A,2}^{\alpha;\beta}\right) \gamma_{A,2}^\beta \exp\left(\pi/4 \gamma_{B,2;A,2}^{\alpha;\beta}\right) +s(\phi) \exp\left(- \pi/4 \gamma_{B,1;A,1}^{\alpha;\beta}\right) \gamma_{A,1}^\beta \exp\left(\pi/4 \gamma_{B,1;A,1}^{\alpha;\beta}\right)\\
&=& -c(\phi) \gamma_{B,2}^\alpha-s(\phi) \gamma_{B,1}^\alpha \\
&=&-\gamma_2 \;, \\
\mathcal{U}^{(6)}(\phi)^\dagger \gamma_2 \mathcal{U}^{(6)}(\phi) &=& \mathcal{U}_{H_1}^{(6)}(\phi)^\dagger \mathcal{U}_{H_2}^{(6)}(\phi)^\dagger \gamma_2 \mathcal{U}_{H_2}^{(6)}(\phi) \mathcal{U}_{H_1}^{(6)}(\phi) \\
&=& \mathcal{U}_{H_1}^{(6)}(\phi)^\dagger \left[c(\phi) \exp\left(- \pi/2 \gamma_{B,2;A,3}^{\alpha;\alpha}\right)\gamma_{B,2}^\alpha \exp\left( \pi/2 \gamma_{B,2;A,3}^{\alpha;\alpha}\right) \right. \\
&& \left.+s(\phi) \exp\left(- \pi/2 \gamma_{B,1;A,2}^{\alpha;\alpha}\right)\gamma_{B,1}^\alpha \exp\left( \pi/2 \gamma_{B,1;A,2}^{\alpha;\alpha}\right)\right] \mathcal{U}_{H_1}^{(1)}(\phi) \\
&=& -  \mathcal{U}_{H_1}^{(6)}(\phi)^\dagger \left[c(\phi)\gamma_{B,2}^\alpha+s(\phi)\gamma_{B,1}^\alpha\right] \mathcal{U}_{H_1}^{(1)}(\phi)^\dagger \\
&=& -c(\phi) \exp\left(- \pi/4 \gamma_{B,2;A,2}^{\alpha;\beta}\right) \gamma_{B,2}^\alpha \exp\left(\pi/4 \gamma_{B,2;A,2}^{\alpha;\beta}\right) -s(\phi) \exp\left(- \pi/4 \gamma_{B,1;A,1}^{\alpha;\beta}\right) \gamma_{B,1}^\alpha \exp\left(\pi/4 \gamma_{B,1;A,1}^{\alpha;\beta}\right)\\
&=& -\gamma_1 \;.
\end{eqnarray*}
	
\end{widetext}

\n Similar to steps 1 and 3, symmetric and antisymmetric superpositions of $\gamma_1$ and $\gamma_2$ form Majorana $\pi$ and zero modes, which are respectively given as $\gamma_\pi^L(\phi)$ and $\gamma_{0,1}^L(\phi)$.

%\n This can be verified by noting that $\exp\left(-\mathrm{i} H_1^{(6)} T/2\right) $ interchanges $(\gamma_{A,2}^\beta,\gamma_{A,1}^\beta)\rightarrow -(\gamma_{B,2}^\alpha,\gamma_{B,1}^\alpha)$ and $(\gamma_{B,2}^\alpha,\gamma_{B,1}^\alpha)\rightarrow (\gamma_{A,2}^\beta,\gamma_{A,1}^\beta)$, whereas $\exp\left(-\mathrm{i} H_2^{(6)} T/2\right) $ flips the sign of $\gamma_{B,2}^\alpha$ and $\gamma_{B,1}^\alpha$.

}
	
\section{Evolution of Majorana modes during $\gamma_{0,2}^L$ and $\gamma_\pi^L$ braiding protocol} \label{app3} 	
	
In the protocol described in Sec.~\ref{prot2}, only $H_2^{(S)}$ is adiabatically deformed, whereas $H_1^{(S)}\equiv H_1$ is kept constant, {  so that the Floquet operator can be written as

 \begin{equation}
 u^{(S)}(\phi)=u_{H_2}^{(S)}(\phi)\times u_{H_1}(\phi) \;,
 \label{fstep2}
 \end{equation}

 \n where $u_{H_2}^{(S)}=\exp\left(-\mathrm{i} H_2^{(S)} T/2\right)$, $u_{H_1}=\exp\left(-\mathrm{i} H_1 T/2\right)$, and $S=1,2,\cdots 7$.} In Majorana basis, $H_2^{(S)}$ can be expressed as (keeping only terms in the first $n$ lattice sites for brevity)

\begin{widetext}
\begin{eqnarray}
H_2^{(1)} T &=& \sum_{k=1}^n \mathrm{i} \pi \left[s(\phi) \gamma_{A,k;B,k}^{\beta;\beta}+c(\phi) \gamma_{B,k;A,k+1}^{\beta;\beta}+\gamma_{B,k;A,k+1}^{\alpha;\alpha} \right] \;, \nonumber \\
H_2^{(2)} T &=& \mathrm{i} \pi\left[\xi_n +c(\phi) \gamma_{B,n;A,n+1}^{\alpha;\alpha}+s(\phi) \gamma_{A,n+1;A,n+1}^{\alpha;\beta}+\gamma_{A,n;B,n}^{\beta;\beta}\right]\;, \nonumber \\
H_2^{(3)} T &=& \mathrm{i} \pi \left[\xi_n +\mathcal{C}(s) \gamma_{A,n+1;A,n+1}^{\alpha;\beta}+\mathcal{C}(s) \gamma_{A,n;B,n}^{\beta;\beta}+\mathcal{S}(s) \left(\gamma_{B,n;A,n+1}^{\beta;\beta}+\gamma_{B,n;A,n+1}^{\alpha;\alpha}\right)\right] \;, \nonumber \\
H_2^{(4)} T &=& \mathrm{i} \pi \left[\xi_n + s(\phi) \gamma_{A,n;B,n}^{\beta;\beta}+c(\phi)\gamma_{B,n;A,n+1}^{\beta;\beta}+\gamma_{B,n;A,n+1}^{\alpha;\alpha}\right]\;, \nonumber \\
H_2^{(5)} T&=& H_2^{(2)} T \;, \nonumber \\
H_2^{(6)} T &=& H_2^{(3)} T \;, \nonumber \\
H_2^{(7)} T &=&  \sum_{k=1}^n \mathrm{i} \pi \left[c(\phi) \gamma_{A,k;B,k}^{\beta;\beta}+s(\phi) \gamma_{B,k;A,k+1}^{\beta;\beta}+\gamma_{B,k;A,k+1}^{\alpha;\alpha} \right] \;,
\end{eqnarray}

\end{widetext}

\noindent where $\xi_n=\sum_{k=1}^{n-1}\left(\gamma_{A,k;B,k}^{\beta;\beta}+\gamma_{B,k;A,k+1}^{\beta;\beta}+\gamma_{B,k;A,k+1}^{\alpha;\alpha}\right)$, $\mathcal{C}=\left(1-f(s)\right)/2$, $\mathcal{S}=\left(1+f(s)\right)/2$, and $f(s)$ is defined in Sec.~\ref{prot2}. Following the same discussion as Appendix~\ref{app2}, {  we will now present the evolution of Majorana modes under the aforementioned adiabatic deformation in steps 1, 2, 4, 5, and 7. As elucidated in Sec.~\ref{prot2}, steps 3 and 6 involve a special two-period adiabatic deformation which is difficult to keep track analytically. That the outcome of these two steps is as intended can be understood from the similarity between the Hamiltonian $H_2^{(3)}$ and $H_2^{(6)}$ (in the Majorana representation) with that studied in our previous work \cite{RG}, as well as from our numerics in Sec.~\ref{prot2}. Finally, note that throughout the steps in this protocol, only $\gamma_\pi^L$ and $\gamma_{0,2}^L$ are affected, while the other Majorana modes stay intact.

\textit{Step 1:}

\begin{eqnarray}
\gamma_\pi^L (\phi) &=& \sum_{k=1}^{n+1} \left(\gamma_{A,k}^\beta +\gamma_{B,k}^\alpha\right) \cos^{n+1-k}\phi \sin^{k-1}\phi \;, \nonumber \\
\gamma_{0,2}^L (\phi) &=& \sum_{k=1}^{n+1} \left(\gamma_{A,k}^\beta -\gamma_{B,k}^\alpha\right) \cos^{n+1-k}\phi \sin^{k-1}\phi \;, \nonumber \\
\end{eqnarray}

\n The above can be verified by first expressing $\gamma_\pi^L (\phi)$ and $\gamma_{0,2}^L (\phi)$ as symmetric and antisymmetric superpositions of two Majorana operators $\gamma_1=\sum_{k=1}^{n+1}\gamma_{A,k}^\beta \cos^{n+1-k}\phi \sin^{k-1}\phi$ and $\gamma_2=\sum_{k=1}^{n+1}\gamma_{B,k}^\alpha \cos^{n+1-k}\phi \sin^{k-1}\phi$, then showing that $u^{(1)}$ transforms $\gamma_1\rightarrow -\gamma_2$ and vice versa. Indeed,

\begin{widetext}

\begin{eqnarray*}
u^{(1)\dagger} \gamma_1 u^{(1)} &=& u_{H_1}^\dagger \gamma_1 u_{H_1} \\
&=&  \sum_{k=1}^{n+1} \exp\left( -\pi/4 \gamma_{B,k;A,k}^{\alpha,\beta}\right) \gamma_{A,k}^\beta \exp\left( \pi/4 \gamma_{B,k;A,k}^{\alpha,\beta}\right) \cos^{n+1-k}\phi \sin^{k-1}\phi \\
&=& -\sum_{k=1}^{n+1}\gamma_{B,k}^\alpha \cos^{n+1-k}\phi \sin^{k-1}\phi \\
&=& -\gamma_2 \;, \\
u^{(1)\dagger} \gamma_2 u^{(1)} &=& u_{H_1}^\dagger \left\lbrace \sum_{k=1}^{n+1} \exp\left( -\pi/2 \gamma_{B,k;A,k+1}^{\alpha,\alpha}\right) \gamma_{B,k}^\alpha \exp\left( \pi/2 \gamma_{B,k;A,k+1}^{\alpha,\alpha}\right) \cos^{n+1-k}\phi \sin^{k-1}\phi \right\rbrace u_{H_1} \\
&=& -\sum_{k=1}^{n+1} \exp\left( -\pi/4 \gamma_{B,k;A,k}^{\alpha,\beta}\right) \gamma_{B,k}^\alpha \exp\left( \pi/4 \gamma_{B,k;A,k}^{\alpha,\beta}\right) \cos^{n+1-k}\phi \sin^{k-1}\phi \\
&=& -\gamma_1 \;,
\end{eqnarray*}
	
\end{widetext}

\n where we have used the fact that $\gamma_1$ commutes with $H_2^{(1)}$ in the above.

%\n This is verified by noting that $\exp\left(-\mathrm{i} H_1 T/2\right) $ interchanges $\gamma_{A,k}^\beta\rightarrow -\gamma_{B,k}^\alpha$ and $\gamma_{B,k}^\beta\rightarrow -\gamma_{A,k}^\alpha$, whereas $\exp\left(-\mathrm{i} H_2^{(1)} T/2\right) $ flips the sign of $\gamma_{B,k}^\alpha$.

\textit{Step 2:}

\begin{eqnarray}
\gamma_\pi^L(\phi) &=& \left[c(\phi)\gamma_{A,n+1}^\beta+s(\phi)\gamma_{B,n}^\alpha\right] \nonumber \\
&& + \left[ c(\phi)\gamma_{B,n+1}^\alpha - s(\phi)\gamma_{A,n}^\beta \right]\;, \nonumber \\
\gamma_{0,2}^L(\phi) &=&\left[c(\phi)\gamma_{A,n+1}^\beta +s(\phi)\gamma_{B,n}^\alpha \right] \nonumber \\
&& - \left[c(\phi)\gamma_{B,n+1}^\alpha - s(\phi)\gamma_{A,n}^\beta \right]\;. \nonumber \\
\end{eqnarray}

\n As before, let $\gamma_1=\left[c(\phi)\gamma_{A,n+1}^\beta+s(\phi)\gamma_{B,n}^\alpha\right]$ and $\gamma_2=\left[ c(\phi)\gamma_{B,n+1}^\alpha - s(\phi)\gamma_{A,n}^\beta \right]$, our objective is to show that $u^{(2)}$ maps $\gamma_1\rightarrow -\gamma_2$.

\begin{widetext}

\begin{eqnarray*}
u^{(2)\dagger} \gamma_1 u^{(2)} &=& u_{H_1}^\dagger \gamma_1 u_{H_1} \\
&=&  c(\phi) \exp\left( -\pi/4 \gamma_{B,n+1;A,n+1}^{\alpha,\beta}\right) \gamma_{A,n+1}^\beta \exp\left( \pi/4 \gamma_{B,n+1;A,n+1}^{\alpha,\beta}\right) +s(\phi) \exp\left( -\pi/4 \gamma_{B,n;A,n}^{\alpha,\beta}\right) \gamma_{B,n}^\alpha \exp\left( \pi/4 \gamma_{B,n;A,n}^{\alpha,\beta}\right)  \\
%&=& -c(\phi) \gamma_{B,n+1}^\alpha +s(\phi)\gamma_{A,n}^\beta \\
&=& -\gamma_2 \;, \\
\end{eqnarray*}

\begin{eqnarray*}
u^{(2)\dagger} \gamma_2 u^{(2)} &=& u_{H_1}^\dagger \left\lbrace c(\phi) \exp\left( -\pi/2 \gamma_{B,n+1;A,n+2}^{\alpha,\alpha}\right) \gamma_{B,n+1}^\alpha \exp\left( \pi/2 \gamma_{B,n+1;A,n+2}^{\alpha,\alpha}\right) \right. \\
&& \left. -s(\phi) \exp\left( -\pi/2 \gamma_{A,n;B,n}^{\beta,\beta}\right) \gamma_{A,n}^\beta \exp\left( \pi/2 \gamma_{A,n;B,n}^{\beta,\beta}\right)\right\rbrace u_{H_1} \\
&=& -c(\phi) \exp\left( -\pi/4 \gamma_{B,n+1;A,n+1}^{\alpha,\beta}\right) \gamma_{B,n+1}^\alpha \exp\left( \pi/4 \gamma_{B,n+1;A,n+1}^{\alpha,\beta}\right) + s(\phi) \exp\left( -\pi/4 \gamma_{B,n;A,n}^{\alpha,\beta}\right) \gamma_{A,n}^\beta \exp\left( \pi/4 \gamma_{B,n;A,n}^{\alpha,\beta}\right) \\
&=& -\gamma_1 \;.
\end{eqnarray*}

\end{widetext}

%\n This is verified by noting that $\exp\left(-\mathrm{i} H_1 T/2\right) $ interchanges $\left(\gamma_{A,n+1}^\beta,\gamma_{A,n}^\beta\right)\rightarrow -\left(\gamma_{B,n+1}^\alpha,\gamma_{B,n}^\alpha\right)$ and $\left(\gamma_{B,n+1}^\alpha,\gamma_{B,n}^\alpha\right)\rightarrow \left(\gamma_{A,n+1}^\beta,\gamma_{A,n}^\beta\right)$, whereas $\exp\left(-\mathrm{i} H_2^{(1)} T/2\right) $ flips the sign of $\gamma_{B,n+1}^\alpha$ and $\gamma_{A,n}^\beta$.

\textit{Step 4:}

\begin{eqnarray}
\gamma_\pi^L(\phi) &=& c(\phi)\gamma_{A,n}^\beta +s(\phi) \gamma_{A,n+1}^\beta \;, \nonumber \\
\gamma_{0,2}^L(\phi) &=& -c(\phi) \gamma_{B,n}^\alpha -s(\phi)\gamma_{B,n+1}^\alpha \;.
\end{eqnarray}

\n Following the end of step 3, $\gamma_\pi^L$ and $\gamma_{0,2}^L$ above are no longer Majorana $\pi$ and zero modes in this step, but they are still zero modes of $(u^{(4)})^2$. These can be directly verified as

\begin{widetext}

\begin{eqnarray*}
u^{(4)\dagger} \gamma_\pi^L u^{(4)} &=& u_{H_1}^\dagger \gamma_\pi^L u_{H_1} \\
&=&  c(\phi) \exp\left( -\pi/4 \gamma_{B,n;A,n}^{\alpha,\beta}\right) \gamma_{A,n}^\beta \exp\left( \pi/4 \gamma_{B,n;A,n}^{\alpha,\beta}\right) +s(\phi) \exp\left( -\pi/4 \gamma_{B,n+1;A,n+1}^{\alpha,\beta}\right) \gamma_{A,n+1}^\beta \exp\left( \pi/4 \gamma_{B,n+1;A,n+1}^{\alpha,\beta}\right)  \\
&=& -c(\phi) \gamma_{B,n}^\alpha -s(\phi)\gamma_{B,n+1}^\alpha \\
&=& \gamma_{0,2}^L \;, \\
u^{(4)\dagger} \gamma_{0,2}^L u^{(4)} &=& -u_{H_1}^\dagger \left\lbrace c(\phi) \exp\left( -\pi/2 \gamma_{B,n;A,n+1}^{\alpha,\alpha}\right) \gamma_{B,n}^\alpha \exp\left( \pi/2 \gamma_{B,n;A,n+1}^{\alpha,\alpha}\right) \right. \\
&& \left. +s(\phi) \exp\left( -\pi/2 \gamma_{B,n+1;A,n+1}^{\alpha,\alpha}\right) \gamma_{B,n+1}^\alpha \exp\left( \pi/2 \gamma_{B,n+1;A,n+1}^{\alpha,\alpha}\right)\right\rbrace u_{H_1} \\
&=& c(\phi) \exp\left( -\pi/4 \gamma_{B,n;A,n}^{\alpha,\beta}\right) \gamma_{B,n}^\alpha \exp\left( \pi/4 \gamma_{B,n;A,n}^{\alpha,\beta}\right) + s(\phi) \exp\left( -\pi/4 \gamma_{B,n+1;A,n+1}^{\alpha,\beta}\right) \gamma_{B,n+1}^\alpha \exp\left( \pi/4 \gamma_{B,n+1;A,n+1}^{\alpha,\beta}\right) \\
&=& \gamma_\pi^L \;.
\end{eqnarray*}

\end{widetext}

\n By combining the two results above, it follows that $\gamma_\pi^L$ and $\gamma_{0,2}^L$ commute with $(u^{(4)})^2$, but they neither commute nor anticommute with $u^{(4)}$.
%In fact, $\gamma_\pi^L=\gamma_\pi+\gamma_0$ and $\gamma_{0,2}^L=\gamma_\pi-\gamma_0$, where $\gamma_\pi$ and $\gamma_0$ are instantaneous Majorana $\pi$ and zero modes of $\mathcal{U}^{(4)}$, which are given by

%\begin{eqnarray}
%\gamma_\pi(\phi) &=& \left[ c(\phi)\gamma_{A,n}^\beta +s(\phi) \gamma_{A,n+1}^\beta \right]+\left[c(\phi)\gamma_{B,n}^\alpha +s(\phi) \gamma_{B,n+1}^\alpha\right] \;, \nonumber \\
%\gamma_0(\phi) &=& \left[ c(\phi)\gamma_{A,n}^\beta +s(\phi) \gamma_{A,n+1}^\beta \right]-\left[c(\phi)\gamma_{B,n}^\alpha +s(\phi) \gamma_{B,n+1}^\alpha\right]\;, \nonumber \\
%\end{eqnarray}

%\n and can be verified in the same way as step 2.

\textit{Step 5:}

\n The Hamiltonian in step 5 evolves in the same way as that in step 2, so the Majorana zero and $\pi$ modes follow those described in step 2. However, continuing step 4, $\gamma_\pi^L$ and $\gamma_{0,2}^L$ are not Majorana $\pi$ and zero modes at this step. Instead, they are given as symmetric and antisymmetric superpositions of Majorana $\pi$ and zero modes found in step 2, so that

\begin{eqnarray}
\gamma_\pi^L(\phi) &=& c(\phi) \gamma_{A,n+1}^\beta +s(\phi) \gamma_{B,n}^\alpha \;, \nonumber \\
\gamma_{0,2}^L(\phi) &=& s(\phi)\gamma_{A,n}^\beta- c(\phi) \gamma_{B,n+1}^\alpha \;.
\end{eqnarray}

\n In particular, these are precisely $\gamma_1$ and $\gamma_2$ defined in step 2, and as shown in that step, $\gamma_\pi^L(\phi)$ and $\gamma_{0,2}^L(\phi)$ indeed commute with $(u^{(5)})^2$.

\textit{Step 7:}

\begin{eqnarray}
\gamma_\pi^L (\phi) &=& \sum_{k=1}^{n+1} \left(\gamma_{A,k}^\beta -\gamma_{B,k}^\alpha\right) \sin^{n+1-k}\phi \cos^{k-1}\phi \;, \nonumber \\
\gamma_{0,2}^L (\phi) &=& -\sum_{k=1}^{n+1} \left(\gamma_{A,k}^\beta +\gamma_{B,k}^\alpha\right) \sin^{n+1-k}\phi \cos^{k-1}\phi \;, \nonumber \\
\end{eqnarray}

\n Note that at this step, $\gamma_\pi^L$ and $\gamma_{0,2}^L$ are Majorana zero and $\pi$ modes respectively, which can be verified by first defining $\gamma_1=\sum_{k=1}^{n+1}\gamma_{A,k}^\beta \sin^{n+1-k}(\phi) \sin^{k-1}(\phi)$ and $\gamma_2=\sum_{k=1}^{n+1}\gamma_{B,k}^\alpha \sin^{n+1-k}(\phi) \cos^{k-1}(\phi)$, then showing that $u^{(7)}$ transforms $\gamma_1\rightarrow -\gamma_2$ and vice versa. Note that $H_2^{(7)}$ is the same as $H_2^{(1)}$ upon taking $c(\phi)\rightarrow s(\phi)$ and $s(\phi)\rightarrow c(\phi)$. As such, the fact that $u^{(7)\dagger} \gamma_{1(2)}u^{(7)}=-\gamma_{2(1)}$ follows exactly the same way as the expansion presented in step 1.

}

{ 
\section{Implementation of our system with cooper pair box} \label{app4}

As outlined in Sec.~\ref{exper}, a possible implementation of our protocol in the proximitized semiconductor setup is through the use of Majorana cooper pair box (MCB) introduced in Ref.~\cite{Tjun2,flux}. The main component of a single MCB consists of a superconducting island, proximitized semiconducting wire accomodating a pair of Majorana modes, and a split Josephson junction enclosing a magnetic flux $\Phi$, as depicted in Fig.~\ref{mcb}(a). In such a setup, the coupling between the two Majorana modes can be varied by tuning $\Phi$.

The Hamiltonian $H_1$ and $H_2$ defined in Eq.~(\ref{H}) can in principle be simulated by designing an array of such MCBs. In particular, since the use of MCB addresses a pair of Majorana modes directly, both real and imaginary hopping or pairing can be realized on equal footing. Indeed, since both $H_1$ and $H_2$ can be recast in terms of Majorana operator bilinears as shown in Eq.~(\ref{Hm}), a possible design of MCB array realizing both $H_1$ and $H_2$ is depicted in Fig.~\ref{mcb}(b) for two lattice sites. There, terms in $H_1$ ($H_2$) are realized by setting $\Phi_{H_2,j}$ ($\Phi_{H_1,j}$) to a value near $\Phi_0/2=\frac{h}{4e}$ (so as to hybridize the respective ancillary Majorana modes) while appropriately setting $\Phi_{H_1,j}$ ($\Phi_{H_2,j}$) to another value which depends on the desired coupling strength \cite{Tjun2,flux}. Periodic quenching between $H_1$ and $H_2$ can then be carried out by periodically quenching the respective fluxes between $\Phi_0/2$ and another value. Such control of magnetic field is expected to be plausible with current technology \cite{magcon}. Finally, $\Phi_{\rm anc, j}$ serve as ancillary fluxes that can be used to accommodate the readout protocol outlined in Sec.~\ref{prot3}. During encoding and braiding processes, these fluxes can simply be switched off.

Finally, we would like to point out that while the minimal design shown in Fig.~\ref{mcb}(b) realizes our original time-periodic Hamiltonian of Eq.~(\ref{model}), it is not sufficient to carry out the braiding protocols described in Sec.~\ref{prot1} and Sec.~\ref{prot2}. For the implementation of these protocols, it is necessary to install additional MCBs into the design to enable coupling between pairs of Majorana operators involved in the steps of our protocols. Although incorporating these additional MCBs may result in an even more complicated design, adiabatic manipulation prescribed in our protocols can be executed by simply tuning the appropriate magnetic fluxes.

\begin{figure*}
	\begin{center}
		\includegraphics[scale=1]{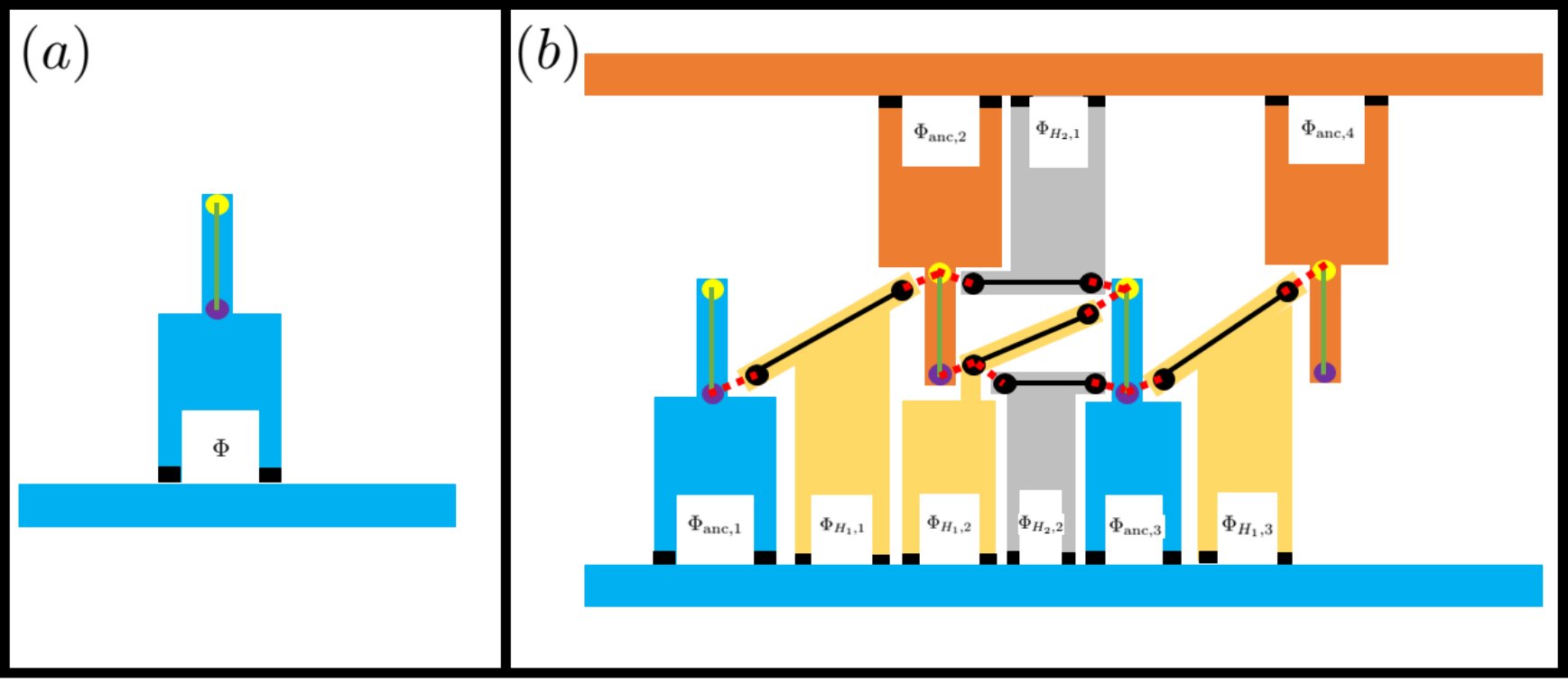}
	\end{center}
	\caption{(color online). (a) Schematic of a single MCB containing a superconducting island (blue), capacitively coupled to a bulk superconductor through a split Josephson junction (black), enclosing a magnetic flux $\Phi$, and a proximitized semiconductor wire (green) hosting a pair of Majorana zero modes (yellow and purple circles). (b) Possible realization of Eq.~(\ref{model}) using an array of MCBs. Blue and brown islands host Majorana operators belonging to sublattice $A$ and $B$ respectively. Yellow and purple circles depict the two species of Majorana operators $\gamma_{A(B),j}^\alpha$ and $\gamma_{A(B),j}^\beta$ respectively in Eq.~(\ref{model}). Black wires and circles depict the ancillary wires with their associated Majorana zero modes to mediate coupling between a pair of Majorana operators $\gamma_{A(B),j}^{\alpha(\beta)}$. Red dotted lines denote tunnel coupling between $\gamma_{A(B),j}^{\alpha(\beta)}$ and ancillary Majorana operators.}
	\label{mcb}
\end{figure*}

}	
	%%%%%%%%%%%%%%%%%%%%%%%%%%%%%% REFERENCE %%%%%%%%%%%%%%%%%%%%%%%%%%%%%%%%%%%%%%%


\begin{thebibliography}{99}
\bibitem{note} Since $\gamma_{0,1}^{L(R)}$ and $\gamma_{0,2}^{L(R)}$ are degenerate, one might argue that the local fermions can be arbitrarily encoded in any superpositions of $\gamma_{0,1}^{L(R)}$ and $\gamma_{0,2}^{L(R)}$, such as $f_0^{L(R)}=a\gamma_{0,1}^{L(R)}+b\gamma_{0,2}^{L(R)}+\mathrm{i}\left(b\gamma_{0,1}^{L(R)}-a\gamma_{0,2}^{L(R)}\right)$. While this is indeed true, it is noted that the qubit encoding and operations presented in this paper do not depend on the specific choice of how one defines the fermion operators since such a choice does not affect the associated parity operators which are used to define our qubits.
\end{thebibliography}
\end{document}